\documentclass[11pt,a4paper]{article}
\usepackage[margin=2.5cm]{geometry}
\usepackage[T1]{fontenc}
\usepackage{lmodern}
\usepackage{amsmath,amssymb,amsthm}
\usepackage{booktabs,graphicx,array}
\usepackage{enumitem}
\usepackage{tikz}
\usetikzlibrary{arrows.meta,positioning,calc,fit,backgrounds,shapes.geometric}
\tikzset{
  blk/.style={draw,rounded corners=1pt,minimum height=7mm,minimum width=12mm,align=center,font=\small},
  sum/.style={draw,circle,inner sep=0pt,minimum size=4mm,font=\scriptsize},
  sig/.style={-{Latex[length=2mm]},thick},
  nd/.style={draw,circle,inner sep=1pt,minimum size=7mm,font=\small},
  ex/.style={draw,rectangle,inner sep=2pt,minimum size=5mm,font=\small,fill=black!5},
  dir/.style={-{Latex[length=2mm]},thick},
  bid/.style={{Latex[length=2mm]}-{Latex[length=2mm]},thick,dashed,bend left=25},
  scc/.style={draw,dashed,rounded corners=4pt,inner sep=4pt,gray},
  lab/.style={font=\scriptsize,fill=white,inner sep=1pt},
}

\usepackage{capt-of}
\usepackage{cite}
\usepackage{placeins}
\usepackage{pgfplots}
\pgfplotsset{compat=1.18}
\newcommand{\ptfig}[3]{\par\medskip\noindent\begin{minipage}{\linewidth}\centering #1\par\smallskip\captionof{figure}{#2}\label{#3}\end{minipage}\par\medskip}

\usepackage[hidelinks]{hyperref}
\hypersetup{pdftitle={Feedback, condensation and induced confounding: a causal framework for dynamical and engineered systems},
pdfauthor={Navdeep M. Singh},pdfsubject={},pdfkeywords={},pdfcreator={LaTeX},pdfproducer={}}
\setlist{itemsep=2pt,topsep=4pt}

\newtheorem{theorem}{Theorem}
\newtheorem{proposition}{Proposition}
\newtheorem{corollary}{Corollary}
\newtheorem{lemma}{Lemma}
\theoremstyle{definition}
\newtheorem{definition}{Definition}
\newtheorem{assumption}{Assumption}
\newtheorem{example}{Example}
\theoremstyle{remark}
\newtheorem{remark}{Remark}

\newcommand{\T}{^{\top}}
\newcommand{\R}{\mathbb{R}}

\title{Feedback, condensation and induced confounding:\\ a causal framework for dynamical and engineered systems}
\author{
Siddhesh Desai\thanks{M.Tech. Student, Department of Electrical Engineering, Veermata Jijabai Technological Institute (VJTI), Mumbai, India, e-mail: skdesai\_m25@ee.vjti.ac.in}
\and
Madhavi Parimi\thanks{Research Scholar, E-MC$^2$RC,   VJTI, Mumbai, India.}
\and
Navdeep Singh\thanks{IGI Research Chair Professor, E-MC$^2$RC, VJTI, Mumbai, India.}
}
\date{}

\begin{document}
\maketitle

\begin{abstract}
\noindent
Structural causal models were built for acyclic systems, but engineered and natural dynamical systems are full of feedback. Cyclic structural causal models give feedback an equilibrium meaning: a cyclic subsystem can be marginalised, or condensed, when its equilibrium equations are uniquely solvable. We call this Regime~2a. It says where a loop settles, not whether it gets there or how it fluctuates. This paper introduces Regime~2b, in which the settled response is also normally attracting. A noisy fast subsystem then leaves bidirected edges among the variables it touches, with stationary weights given by the Lyapunov equation in continuous time and the Stein equation in discrete time; sampling maps one exactly onto the other. Regime~2b implies Regime~2a but not conversely: at a Hopf, flip or Neimark--Sacker crossing the condensed graph is unchanged while the edge weights diverge. The confounding gain $\|\Pi\|=\eta/(2|\alpha|)$ is the exact worst-case amplification from a fluctuation--dissipation defect to induced confounding, and it splits into critical slowing down and non-normal amplification. We compute equilibrium causal orderings by bipartite matching and show that controllers with integral action are sinks of that ordering, so their gains shape the fluctuations but not the settled state. We state what interventions mean in each regime, show that least squares on closed-loop data can get the sign of an effect wrong while an instrument recovers it, and give the time-scale separation a slower layer needs before it may treat a faster one as condensed. Worked examples, all checked numerically, and two engineering applications illustrate the results. The paper is intended as the reference statement of the framework.
\end{abstract}

\section{Orientation}\label{sec:orientation}

Picture a power converter connected to the grid, with two meters at its terminals: one for voltage, one for current. Both readings jitter. Some of the jitter comes from the grid, and some from inside the converter: switching ripple and sensor noise stir its current controller and its phase-locked loop, and those loops push on both meters at once. Anyone who reads only the meters sees two signals that move together for no visible reason. In causal language, the converter's internal loops are a hidden common cause, and they leave a bidirected edge between the two readings.

The situation is common. A building's heating supervisor reads valve positions that the room loops move together. A traffic supervisor reads spacing errors that a platoon's leader moves together. A power dispatcher reads power flows that the converters' fast controls move together. In each case there is a fast feedback subsystem, a slower layer or an analyst who reads its outputs, and a question: what can be concluded from those outputs, and what cannot?

Causal reasoning about such systems needs two things. The first is to know what the fast subsystem does on average: where it settles, and how that settling point depends on the slower variables. This is a statement about equilibria. If the fast subsystem's equilibrium is unique, it can be replaced by its equilibrium map, and the slower variables then act on the outputs through that map. We call the replacement \emph{condensation}, and the setting in which it is possible \emph{Regime~2a}. Its rigorous foundation is the theory of cyclic structural causal models \cite{bongers2021}.

The second is to know what the fast subsystem does around its average. Noise does not stop at the equilibrium. It is filtered by the fast dynamics and reaches every output the subsystem touches, so the outputs become correlated through the subsystem. How strongly depends on how quickly the subsystem forgets a disturbance and on how it routes disturbances between its states. We call the setting in which these correlations are stationary and finite \emph{Regime~2b}. It requires the fast subsystem to be stable, which asks more than a unique equilibrium.

A two-state example shows why the second requirement is separate from the first. Take
\begin{equation}\label{eq:two1}
\dot x=\begin{pmatrix}-d&a\\ b&-d\end{pmatrix}x+\dot w,
\end{equation}
with $w$ a pair of independent Wiener processes. Its stationary covariance is
\begin{equation}\label{eq:two2}
P_{11}=\frac{a^2-ab+2d^2}{4d(d^2-ab)},\qquad P_{22}=\frac{b^2-ab+2d^2}{4d(d^2-ab)},\qquad P_{12}=\frac{a+b}{4(d^2-ab)} .
\end{equation}
Three things can be read off. If $b=-a$, the coupling only exchanges energy between the states and $P_{12}=0$: the loop induces no correlation at all. If $ab\to d^2$, an eigenvalue moves to zero, the equilibrium stops being unique, and $P_{12}$ diverges; condensation fails together with everything else. But if $ab<0$ and the damping $d$ goes to zero, the eigenvalues are $-d\pm j\sqrt{-ab}$ and cross the imaginary axis away from the origin. The determinant stays near $-ab$, the equilibrium stays unique and the condensed map is unchanged. Yet the variances grow like $1/d$, the cross-spectrum at the oscillation frequency grows like $1/d^2$, and the induced edges between outputs that share the oscillation blow up. A model that knows only the equilibrium cannot see this happening.

\subsection{The two regimes in brief}
\emph{Regime 2a} concerns where a feedback loop settles. Take a room heated by a valve under a thermostat. Whatever the outside temperature, the loop drives the room to its setpoint, and the valve opening at which it settles is fixed by the heat balance of the room. Someone who only cares about the long run can replace the whole loop by that heat balance: tell it the outside temperature and the setpoint, and it returns the valve opening. That replacement is condensation, and it is legitimate when the settling point is unique. Nothing about gains, sampling or noise enters it. In fact, with integral action in the thermostat the settling point does not even depend on the controller's tuning (Section~\ref{sec:2a}).

\emph{Regime 2b} concerns how the loop moves around that settling point. Gusts at the window and noise in the temperature sensor keep the valve and the room temperature moving, and they move together, because the loop links them. How much they move, and how strongly together, depends on how fast the loop forgets a disturbance and how it passes disturbances between its states. Over-tune a digital thermostat and its valve starts to alternate from one sample to the next. The settling point is unchanged, so the condensed model sees nothing, but the valve and the temperature now swing together with growing amplitude. Regime~2b is the setting in which these joint fluctuations are stationary and finite. Their size is computed from the Lyapunov equation in continuous time, or the Stein equation in discrete time, and Section~\ref{sec:2b} bounds it.

The first regime is what cyclic causal models already capture. The second is the subject of this paper.

\subsection{Outline}
This paper sets out both regimes in one framework. Section~\ref{sec:background} recalls what is known about causal models with feedback and with dynamics, and says where this work fits. Section~\ref{sec:setting} fixes the setting in continuous and discrete time. Section~\ref{sec:regime2} introduces Regime~2, the setting of feedback cycles in causal models, and defines its two parts: Regime~2a, from existing work, and Regime~2b, from this paper. Section~\ref{sec:2a} develops Regime~2a, including how the equilibrium causal ordering is computed for plants with one state or many, and Section~\ref{sec:2b}, the main part, develops Regime~2b: Lyapunov and Stein equations, frequency-resolved edges, the exits from the regime, the confounding gain and its two factors, the fluctuation--dissipation defect, interventions, and identification. Section~\ref{sec:separation} gives conditions under which a slower layer may treat a faster one as condensed. Section~\ref{sec:naim} places both regimes in the geometry of normally attracting, asymptotically invariant manifolds, Section~\ref{sec:points} lists point by point what attraction adds to solvability, and Section~\ref{sec:offers} says in plain terms what the formulation offers a designer. Section~\ref{sec:examples} works five examples and summarises two applications. Section~\ref{sec:use} collects the framework into a checklist, and Section~\ref{sec:scope} states its limits.

\section{Causal models, feedback and dynamics}\label{sec:background}

\subsection{Structural causal models and confounding}
A structural causal model assigns each endogenous variable a function of its parents and of exogenous noise \cite{pearl2009,spirtes2000}. Its graph has a directed edge from each parent to its child. When two variables share an unobserved common cause, marginalising that cause leaves a bidirected edge between them, and the resulting acyclic directed mixed graph (ADMG) records both kinds of dependence. An intervention $\mathrm{do}(X=x)$ replaces the equation of $X$ by the constant $x$ and deletes the edges into $X$. Confounding is the gap between what conditioning on $X$ predicts and what intervening on $X$ does. Bidirected edges are its graphical trace, and much of causal identification is about routes around them: adjustment sets, front-door paths, and instrumental variables.

\subsection{Feedback and equilibrium}
Acyclicity rules out feedback, so the classical theory does not apply directly to engineered systems. Two lines of work bring feedback in through equilibria. The older is Simon's causal ordering \cite{simon1953}, which reads causal structure off a set of equilibrium equations by finding self-contained subsets; in modern form this is the Dulmage--Mendelsohn decomposition of the equation--variable incidence graph \cite{dulmage1958,blom2021}. Iwasaki and Simon discussed abstracting a model by aggregating subsystems that equilibrate quickly \cite{iwasaki1994}, which is close in spirit to condensation. The more recent line is the theory of cyclic structural causal models with latent variables \cite{spirtes1995,bongers2021,forre2017}. Bongers, Forr\'e, Peters and Mooij give such models a rigorous foundation, including the conditions under which a cyclic subsystem can be marginalised: the subsystem's equations must be uniquely solvable for its own variables given the rest \cite{bongers2021}. Their semantics are equilibrium semantics. A cyclic structural causal model describes where a feedback system settles, not how it gets there or how it fluctuates.

Equilibrium models have known pitfalls. An equilibrium model need not commute with manipulation of the dynamics that produced it \cite{dash2005}, and the equilibrium of a manipulated system exists and is reached only under conditions on the dynamics \cite{mooij2013}. Both points matter below.

\subsection{Dynamics and causal models}
A separate body of work connects causal models directly to dynamics. Mooij, Janzing and Sch\"olkopf showed how equilibria of ordinary differential equations give structural causal models \cite{mooij2013}, and Bongers, Blom and Mooij extended this to random differential equations and to stochastic processes \cite{bongers2018}. Boeken and Mooij study dynamic structural causal models whose variables are functions of time, possibly cyclic and with latent confounding \cite{boeken2024}. Sokol and Hansen defined interventions in stochastic differential equations \cite{sokol2014}, and local independence graphs describe dependence between continuous-time processes \cite{mogensen2018}. Closest to the present work are graphical continuous Lyapunov models \cite{fitch2019,varando2020,dettling2023}. They treat each observation as a snapshot of a stationary diffusion whose covariance solves a continuous Lyapunov equation, with a sparse drift matrix encoding causal structure, and they study identifiability and estimation of that drift matrix.

\subsection{Where this paper fits}
The framework here sits between these lines. Like cyclic structural causal models, it marginalises feedback subsystems, and Regime~2a is exactly the condition for doing so. Like the Lyapunov-model literature, it uses the stationary covariance of a linear stochastic system. Its aim differs from both: it asks what a feedback subsystem does to the variables around it, and how that effect can be bounded and designed, rather than how to learn a graph from data. Four things are specific to it:

\begin{enumerate}[label=(\roman*)]
\item the distinction between Regime~2a and Regime~2b as a structural fact, with the exits from 2b at which the condensed graph is unchanged while its bidirected edges diverge;
\item a worst-case law for the weights of those edges, $\|\Pi\|=\eta/(2|\alpha|)$, which separates critical slowing down from non-normal amplification and ties both to a fluctuation--dissipation defect;
\item condensation under finite time-scale separation, with frequency-resolved edges and explicit error terms for what a slower layer actually sees;
\item the use of all of this in design: controllers as sinks of the causal ordering, gain-free condensed maps under the internal-model principle, and interfaces between layers of a control architecture \cite{singh2026c}.
\end{enumerate}

\section{Setting}\label{sec:setting}

\subsection{Continuous time}
A \emph{fast block} with state $x_f\in\R^n$ is driven by a slower variable $x_s$ and by noise, and it produces outputs $y$:
\begin{equation}\label{eq:ct}
\dot x_f=Ax_f+Hx_s+B\dot w,\qquad y=Cx_f+D_px_s+D_w\dot w+n .
\end{equation}
Here $w$ is a standard Wiener process, $D_w$ carries noise that reaches the outputs directly, and $n$ is measurement noise of intensity $\Sigma_s$, independent of $w$. A slower layer, when one is present, updates $x_s$ from $y$; Section~\ref{sec:separation} treats that coupling. The block may be a linearisation of a nonlinear system about an operating point. Section~\ref{sec:scope} says what carries over to the nonlinear case.

\subsection{Discrete time}
The discrete-time counterpart is
\begin{equation}\label{eq:dt}
x_f(k+1)=A_dx_f(k)+H_dx_s(k)+w(k),\qquad y(k)=Cx_f(k)+D_px_s(k)+n(k),
\end{equation}
with $w(k)$ independent, zero-mean and of covariance $Q_d$. This covers digital controllers, sampled measurements, and systems that are discrete by nature.

\subsection{Coordinates}
Several quantities below depend on coordinates. A \emph{storage metric} is a matrix $Q\succ0$ fixed in advance, normally from stored energy. In energy coordinates $\tilde x=Q^{1/2}x$ we write $\tilde A=Q^{1/2}AQ^{-1/2}$, $\tilde B=Q^{1/2}B$, and similarly in discrete time. Statements that depend on $Q$ say so; quantities defined on the outputs, such as covariances of $y$, do not depend on it.

\section{Regime 2: feedback cycles in causal models}\label{sec:regime2}
We use \emph{Regime 2} for the setting in which a causal model contains a feedback cycle that is to be marginalised, so that the variables outside the cycle can be reasoned about with the cycle collapsed. (Regime~1, the acyclic case of Pearl's theory, needs no such step.) Regime~2 splits into two parts. Regime~2a asks whether the cycle has a unique settled response, and it is the setting of existing work on cyclic causal models. Regime~2b asks, in addition, whether that settled response attracts the motion of the cycle, and it is the contribution of this paper. This section explains why the split is needed, states both definitions side by side, and says how they relate.

\subsection{From acyclic graphs to feedback: why two regimes}\label{sec:why}
Pearl's causal framework, and most of causal inference built on it, rests on directed acyclic graphs \cite{pearl2009}. Engineered and physical systems are full of feedback, and feedback puts cycles into the causal graph. The usual way to keep the acyclic machinery is to collapse each cycle into a single node. Two operations must be kept apart here. Collapsing the strongly connected components of a graph is purely structural: it can always be done, and it says nothing about what the collapsed node computes. Marginalising a cyclic subsystem, in the sense of cyclic structural causal models \cite{bongers2021}, is stronger. It replaces the subsystem by the map from its inputs to its settled outputs, and this is legitimate when the subsystem's equilibrium equations have a unique solution. That condition, together with the map $J_s$ it produces, is what we call Regime~2a, and the replacement is condensation.

Condensation in this sense is an equilibrium notion. It says where the loop settles, but not whether it gets there, how fast, or how it fluctuates along the way. Dynamics have been brought into causal modelling before, through equilibria of differential equations, dynamic structural causal models and graphical Lyapunov models (Section~\ref{sec:background}). None of these attaches to the condensed graph the effect that a noisy, stable feedback loop has on the variables around it. Keeping only the equilibrium loses three things: whether the loop actually settles, the statistical coupling it induces among the variables it touches, and how much time-scale separation a slower layer needs before it may treat the loop as settled.

Regime~2b recovers these. It does not replace condensation. It keeps the same collapsed node and the same directed edges $J_s$, and asks in addition that the loop be asymptotically stable. The loop then has a stationary covariance, given by the Lyapunov equation \eqref{eq:lyap} in continuous time and the Stein equation \eqref{eq:stein} in discrete time, and its effect on the surrounding variables appears as bidirected edges with computable weights (Figure~\ref{fig:scheme}(c)). The condensed graph becomes an acyclic directed mixed graph, with directed edges that come from the equilibrium and bidirected edges that come from the dynamics. Regime~2b is strictly smaller than Regime~2a, and Section~\ref{sec:points} sets out what separates them. The two regimes are defined in Sections~\ref{sec:def2a} and~\ref{sec:def2b} below and developed in Sections~\ref{sec:2a} and~\ref{sec:2b}.

\subsection{Regime 2a: unique solvability (existing work)}\label{sec:def2a}
Regime~2a is not new. The condition is the one under which cyclic structural causal models can be marginalised \cite{bongers2021}, under which equilibria of differential equations define structural causal models \cite{mooij2013}, and under which Simon's causal ordering assigns every variable to a unique equation \cite{simon1953,blom2021}. What we add is the name, so that it can be set against Regime~2b, and its statement for the linear fast blocks of Section~\ref{sec:setting}.

\begin{definition}[Regime 2a]\label{def:2a}
The continuous-time fast block \eqref{eq:ct} is in Regime~2a if $A$ is nonsingular. The discrete-time block \eqref{eq:dt} is in Regime~2a if $I-A_d$ is nonsingular.
\end{definition}

In Regime~2a the noise-free block has, for each constant $x_s$, the unique equilibrium
\[
x_f^\star=-A^{-1}Hx_s\quad\text{(continuous)},\qquad x_f^\star=(I-A_d)^{-1}H_dx_s\quad\text{(discrete)}.
\]
\emph{Condensation} replaces the block by this map. The outputs then depend on $x_s$ through the condensed sensitivity
\[
J_s=D_p-CA^{-1}H\quad\text{or}\quad J_s=D_p+C(I-A_d)^{-1}H_d .
\]
In the language of cyclic structural causal models, Regime~2a is the unique-solvability condition for the fast block's equilibrium equations with respect to its own variables, which is what permits marginalising it \cite{bongers2021}. The condensed graph keeps a directed edge from $x_s$ to output $y_i$ whenever $(J_s)_i\ne0$. Regime~2a is about where the cycle settles. It does not ask whether the cycle gets there.

\subsection{Regime 2b: normal attraction (this paper)}\label{sec:def2b}
Regime~2b is the dynamic extension that this paper introduces. It keeps the settled response of Regime~2a and asks that it attract.

\begin{definition}[Regime 2b]\label{def:2b}
The continuous-time block is in Regime~2b if $A$ is Hurwitz. The discrete-time block is in Regime~2b if $A_d$ is Schur, that is, its spectral radius $\rho(A_d)$ is less than one.
\end{definition}

The definition comes from a geometric reading. In the language of Normally Attracting, Invariant Manifolds (NAIMs) \cite{naim2026,fenichel1979}, the settled response $x_f=h(x_s)$, here $h(x_s)=-A^{-1}Hx_s$, is a manifold over the slow variables. Regime~2a says that this manifold exists and is unique. Regime~2b says that it is normally attracting: states knocked off it return to it. Section~\ref{sec:naim} develops this correspondence. Normal attraction has two consequences that Regime~2a lacks. The cycle actually settles, at a rate set by the spectral abscissa $\alpha$ of $A$ (or the spectral radius $\rho$ of $A_d$). And when the cycle is driven by noise, its fluctuation about the manifold has a stationary covariance, given by the Lyapunov equation \eqref{eq:lyap} in continuous time and the Stein equation \eqref{eq:stein} in discrete time (Section~\ref{sec:2b}). That covariance gives the bidirected edges of the condensed graph.

\subsection{How the two regimes relate}
The two conditions are not symmetric: one implies the other, but not the reverse.

\begin{proposition}\label{prop:contain}
Regime~2b implies Regime~2a. Regime~2a does not imply Regime~2b.
\end{proposition}

\begin{proof}
A Hurwitz $A$ has every eigenvalue in the open left half-plane, so none is at $0$ and $A$ is nonsingular. A Schur $A_d$ has every eigenvalue in the open unit disk, so none is at $1$ and $I-A_d$ is nonsingular. For the converse it is enough to give blocks that are in Regime~2a and not in Regime~2b. The two-state loop \eqref{eq:two1} with $a=-2$, $b=1$ and $d=0$ has eigenvalues $\pm j\sqrt2$ and $\det A=2$, a Hopf point. The block $A=\operatorname{diag}(-1,2)$ is a saddle with $\det A=-2$. In discrete time, $A_d=-1$ (a flip) gives $I-A_d=2$, and any $A_d$ with a complex pair on the unit circle away from $1$ (a Neimark--Sacker point) also keeps $I-A_d$ nonsingular.
\end{proof}

In each of these examples the equilibrium is unique, so the condensed map $J_s$ and the condensed graph exist, but the block does not settle and has no stationary covariance. Regime~2a fixes where a cycle would settle; only Regime~2b says that it does. The containment is strict, and the exits from Regime~2b that keep Regime~2a are exactly the Hopf, flip and Neimark--Sacker crossings (Section~\ref{sec:exits}); along them the condensed graph is unchanged while its bidirected edges diverge.

\begin{remark}[Nonlinear blocks]\label{rem:local}
For the linear blocks of this paper Proposition~\ref{prop:contain} is exact. For a nonlinear block, Regime~2b is a local statement: normal attraction at an equilibrium makes the normal linearisation nonsingular, and the implicit-function theorem then gives a unique settled response near that equilibrium. Unique solvability in the sense of \cite{bongers2021} asks for a unique solution of the equilibrium equations as a whole. A bistable loop has two equilibria for the same inputs, each normally attracting, so it satisfies Regime~2b locally at both while failing global unique solvability. In the nonlinear setting Regime~2b therefore implies Regime~2a only locally, around the equilibrium in question (Sections~\ref{sec:naim} and~\ref{sec:scope}).
\end{remark}

Table~\ref{tab:regimes} sets the two regimes side by side.

\begin{table}[!htbp]
\centering
\small
\caption{The two parts of Regime~2.}\label{tab:regimes}
\begin{tabular}{@{}p{3.2cm}p{5.4cm}p{5.4cm}@{}}
\toprule
 & Regime 2a & Regime 2b \\
\midrule
Condition & $A$ nonsingular ($I-A_d$ nonsingular) & $A$ Hurwitz ($A_d$ Schur) \\
Question answered & Where does the cycle settle? & Does it settle, and how does it fluctuate about the settled state? \\
Geometric reading & settled-response manifold exists and is unique & that manifold is normally attracting \\
Condensed graph & directed edges $J_s$ & directed edges $J_s$ and bidirected edges $\kappa_{ij}$, $\Phi_{ij}(\omega)$ \\
Main equations & $J_s=D_p-CA^{-1}H$ & Lyapunov and Stein equations \eqref{eq:lyap}, \eqref{eq:stein} \\
Source & Simon \cite{simon1953}; Mooij et al.~\cite{mooij2013}; Bongers et al.~\cite{bongers2021} & this paper \\
\bottomrule
\end{tabular}
\end{table}

\FloatBarrier
\section{Regime 2a: condensation and causal ordering}\label{sec:2a}
This section develops Regime~2a for the fast blocks of Section~\ref{sec:setting}: the two graphs that a feedback model carries, the causal ordering of its equilibrium, and the special place of controllers in that ordering.

\subsection{Two graphs from one model}\label{sec:twographs}
Two graphs are read off the same linear model, and both are used throughout.

The \emph{dynamic causal graph} has a node per state and an edge $x_j\to x_i$ whenever $\partial\dot x_i/\partial x_j\ne0$ (or $(A_d)_{ij}\ne0$). Feedback makes it cyclic. Collapsing its strongly connected components, the loops, leaves a directed acyclic graph of components.

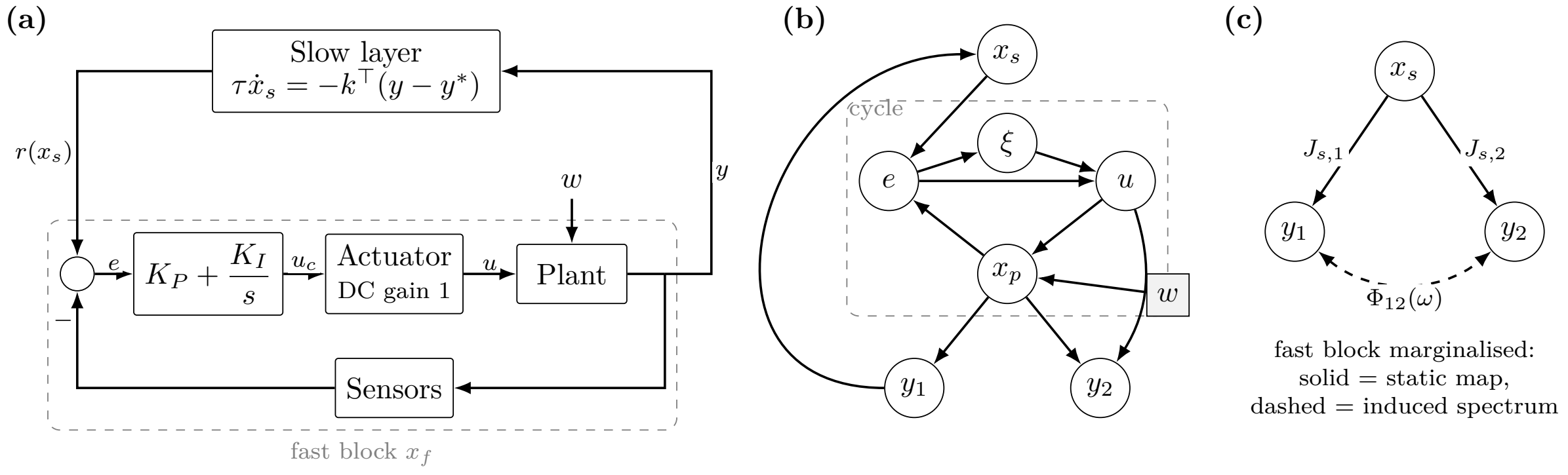
\begin{figure}[!htbp]
\centering
\resizebox{\textwidth}{!}{\begin{tikzpicture}
\begin{scope}
\node[font=\small\bfseries] at (-0.6,3.0) {(a)};
\node[blk,minimum width=34mm] (slow) at (3.3,2.4) {Slow layer\\[-1pt]$\tau\dot x_s=-k^{\top}(y-y^\ast)$};
\node[sum] (s1) at (0,0) {};
\node[blk,minimum width=15mm] (ctrl) at (1.55,0) {$K_P+\dfrac{K_I}{s}$};
\node[blk,minimum width=15mm] (act) at (3.75,0) {Actuator\\[-1pt]{\scriptsize DC gain 1}};
\node[blk,minimum width=13mm] (plant) at (5.85,0) {Plant};
\node[blk] (sens) at (3.75,-1.35) {Sensors};
\coordinate (tap) at (6.95,0);
\draw[sig] (slow.west) -| node[lab,pos=0.72,left]{$r(x_s)$} (s1.north);
\draw[sig] (s1) -- node[lab,above]{$e$} (ctrl);
\draw[sig] (ctrl) -- node[lab,above]{$u_c$} (act);
\draw[sig] (act) -- node[lab,above]{$u$} (plant);
\draw[thick] (plant.east) -- (tap);
\draw[sig] (tap) |- (sens.east);
\draw[sig] (sens.west) -| node[lab,pos=0.85,left]{$-$} (s1.south);
\node[font=\small] (w) at (5.85,1.1) {$w$};
\draw[sig] (w) -- (plant.north);
\draw[sig] (tap) -- ++(0.55,0) |- node[lab,pos=0.25,right]{$y$} (slow.east);
\begin{scope}[on background layer]
\node[scc,fit=(s1)(ctrl)(act)(plant)(sens)(tap),label={[font=\scriptsize,gray]below:fast block $x_f$}] {};
\end{scope}
\end{scope}
\begin{scope}[xshift=9.6cm]
\node[font=\small\bfseries] at (-1.0,3.0) {(b)};
\node[nd] (xs) at (1.4,2.6) {$x_s$};
\node[nd] (e) at (0,1.1) {$e$};
\node[nd] (xi) at (1.4,1.55) {$\xi$};
\node[nd] (u) at (2.8,1.1) {$u$};
\node[nd] (x) at (1.4,0.0) {$x_p$};
\node[ex] (wn) at (3.3,-0.25) {$w$};
\node[nd] (y1) at (0.3,-1.35) {$y_1$};
\node[nd] (y2) at (2.5,-1.35) {$y_2$};
\draw[dir] (xs) -- (e); \draw[dir] (e) -- (xi); \draw[dir] (xi) -- (u); \draw[dir] (e) -- (u);
\draw[dir] (u) -- (x); \draw[dir] (x) -- (e); \draw[dir] (wn) -- (x);
\draw[dir] (x) -- (y1); \draw[dir] (x) -- (y2); \draw[dir] (u) to[bend left=25] (y2);
\draw[dir] (y1) to[out=180,in=180,looseness=1.6] (xs);
\begin{scope}[on background layer]
\node[scc,fit=(e)(xi)(u)(x)] (cyc) {};
\end{scope}
\node[font=\scriptsize,gray] at (-0.15,1.95) {cycle};
\end{scope}
\begin{scope}[xshift=14.4cm]
\node[font=\small\bfseries] at (-0.6,3.0) {(c)};
\node[nd] (xs2) at (1.3,2.4) {$x_s$};
\node[nd] (Y1) at (0,0.5) {$y_1$};
\node[nd] (Y2) at (2.6,0.5) {$y_2$};
\draw[dir] (xs2) -- node[lab,left]{$J_{s,1}$} (Y1);
\draw[dir] (xs2) -- node[lab,right]{$J_{s,2}$} (Y2);
\draw[{Latex[length=2mm]}-{Latex[length=2mm]},thick,dashed] (Y1) to[bend right=40] node[lab,below]{$\Phi_{12}(\omega)$} (Y2);
\node[font=\scriptsize,align=center] at (1.3,-1.25) {fast block marginalised:\\ solid = static map,\\ dashed = induced spectrum};
\end{scope}
\end{tikzpicture}}
\caption{A slow layer over an integrating fast loop. (a) Block diagram. (b) Dynamic causal graph: the fast loop is one cycle (dashed box). (c) The slow layer's view after condensation: directed edges carry the condensed sensitivity $J_s$, the dashed bidirected edge carries the induced cross-spectrum of Regime~2b.}
\label{fig:scheme}
\end{figure}

The \emph{equilibrium causal ordering} describes what determines what once the block has settled \cite{simon1953}. Each equilibrium equation is matched to the variable it determines, using a perfect matching in the equation--variable incidence graph, and an edge $x_k\to x_j$ is drawn when the equation matched to $x_j$ contains $x_k$. The strongly connected components of this graph, in topological order, form the causal ordering; this is the Dulmage--Mendelsohn block-triangular form \cite{dulmage1958,blom2021}. Section~\ref{sec:matching} sets out the computation step by step. Figure~\ref{fig:scheme} shows both graphs for a single integrating loop under a slower layer, together with the condensed view that the slower layer has.

\subsection{Controllers are sinks of the causal ordering}
In engineered feedback, the fast block usually contains integral action or, more generally, an internal model of the signals it must track or reject \cite{francis1976}. This gives Regime~2a a specific and useful structure.

\begin{assumption}\label{ass:im}
(a) The block contains integrators $\dot\xi=e$ on $p$ regulated errors $e=r(x_s)-C_ex_p-n_e$, where $x_p$ is the plant state. (b) The applied input $u\in\R^p$ comes from a stable actuator with unit gain at zero frequency, driven by $u_c=K_I\xi+\Gamma$, where $K_I$ is invertible and $\Gamma$ is a linear function of plant states, measurements, noise and other controller states. (c) Those other controller states obey $\dot\zeta=A_c\zeta+B_ce$ with $A_c$ nonsingular. (d) The matrix $\left(\begin{smallmatrix}A_p&B_p\\ C_e&0\end{smallmatrix}\right)$ is nonsingular. (e) The outputs are functions of plant signals, of $u_c$, of $x_s$ and of noise.
\end{assumption}

\begin{proposition}\label{prop:gainfree}
Under Assumption~\ref{ass:im} the block is in Regime~2a, and $J_s$ and the zero-frequency transfer $T_0=D_w-CA^{-1}B$ from noise to outputs do not depend on the gains in $K_I$, $\Gamma$, $A_c$ or $B_c$.
\end{proposition}

\begin{proof}
Hold $x_s$ and the noise at constant values. The integrator rows give $e=0$, so $C_ex_p=r(x_s)-n_e$; with $A_c$ nonsingular, $\zeta=0$; the actuator gives $u=u_c$. The plant rows and the regulation condition form a linear system in $(x_p,u)$ whose matrix is the one in (d), so $(x_p,u)$ is unique and involves no controller gain. The command law then fixes $\xi=K_I^{-1}(u-\Gamma)$. This is the only place the gains enter, and by (e) $\xi$ does not reach the outputs. The equilibrium outputs are therefore a gain-free function of $(x_s,\bar w)$, where $\bar w$ is the constant noise level. Since the whole equilibrium state exists and is unique for every constant input, $A$ is nonsingular, and $J_s$ and $T_0$ are the derivatives of that function.
\end{proof}

\begin{corollary}\label{cor:regulated}
An output equal to a measured regulated signal $C_ex_p+n_e$ equals $r(x_s)$ at zero frequency. It carries no information about the plant or the disturbances there.
\end{corollary}

\begin{figure}[!htbp]
\centering
\resizebox{\textwidth}{!}{\begin{tikzpicture}
\begin{scope}
\node[font=\small\bfseries] at (-0.5,1.5) {(a)};
\node (r) at (-0.4,0) {$r$};
\node[sum] (s) at (0.5,0) {};
\node[blk,minimum width=15mm] (pi) at (2.05,0) {$K_P+\dfrac{K_I}{s}$};
\node[blk] (ac) at (4.15,0) {$\dfrac{1}{T s+1}$};
\node[blk] (pl) at (6.55,0) {$\dot x_p=A_px_p$\\[-1pt]$+B_pu+E_pd$};
\node (d) at (6.55,1.3) {$d$};
\node[sum] (sn) at (4.15,-1.25) {};
\node (n) at (4.15,-2.05) {$n_e$};
\draw[sig] (r) -- (s); \draw[sig] (s) -- node[lab,above]{$e$} (pi); \draw[sig] (pi) -- node[lab,above]{$u_c$} (ac);
\draw[sig] (ac) -- node[lab,above]{$u$} (pl); \draw[sig] (d) -- (pl);
\draw[sig] (pl.east) -- ++(0.35,0) |- node[lab,pos=0.7,below]{$C_ex_p$} (sn);
\draw[sig] (n) -- (sn); \draw[sig] (sn) -| node[lab,pos=0.88,left]{$-$} (s);
\end{scope}
\begin{scope}[xshift=9.4cm]
\node[font=\small\bfseries] at (-0.6,1.5) {(b)};
\node[nd] (x) at (0,0) {$x_p$};
\node[nd] (e) at (1.4,0.95) {$e$};
\node[nd] (xi) at (2.9,0.95) {$\xi$};
\node[nd] (u) at (2.15,-0.65) {$u$};
\draw[dir] (x) -- (e); \draw[dir] (e) -- (xi); \draw[dir] (xi) -- (u); \draw[dir] (e) -- (u); \draw[dir] (u) -- (x);
\node[font=\scriptsize,align=center] at (1.45,-1.65) {dynamics: one cycle};
\end{scope}
\begin{scope}[xshift=13.9cm]
\node[font=\small\bfseries] at (-0.6,1.5) {(c)};
\node[ex] (r2) at (0,0.95) {$r,n_e$};
\node[ex] (d2) at (0,-0.65) {$d$};
\node[nd] (x2) at (1.45,0.95) {$x_p$};
\node[nd] (u2) at (1.45,-0.65) {$u$};
\node[nd,fill=black!8] (xi2) at (3.1,0.15) {$\xi$};
\draw[dir] (r2) -- (x2); \draw[dir] (x2) -- (u2); \draw[dir] (d2) -- (u2);
\draw[dir] (x2) -- (xi2); \draw[dir] (u2) -- node[lab,below right]{$K_I^{-1}$} (xi2);
\node[font=\scriptsize,align=center] at (1.5,-1.65) {equilibrium: gains only\\ on edges into the sink $\xi$};
\end{scope}
\end{tikzpicture}}
\caption{A PI loop. (a) Block diagram. (b) Dynamic causal graph: one cycle. (c) Equilibrium causal ordering: $x_p$ and $u$ are fixed by exogenous inputs, and the controller state $\xi$ is a sink; the gains enter only on the edges into $\xi$.}
\label{fig:pi}
\end{figure}
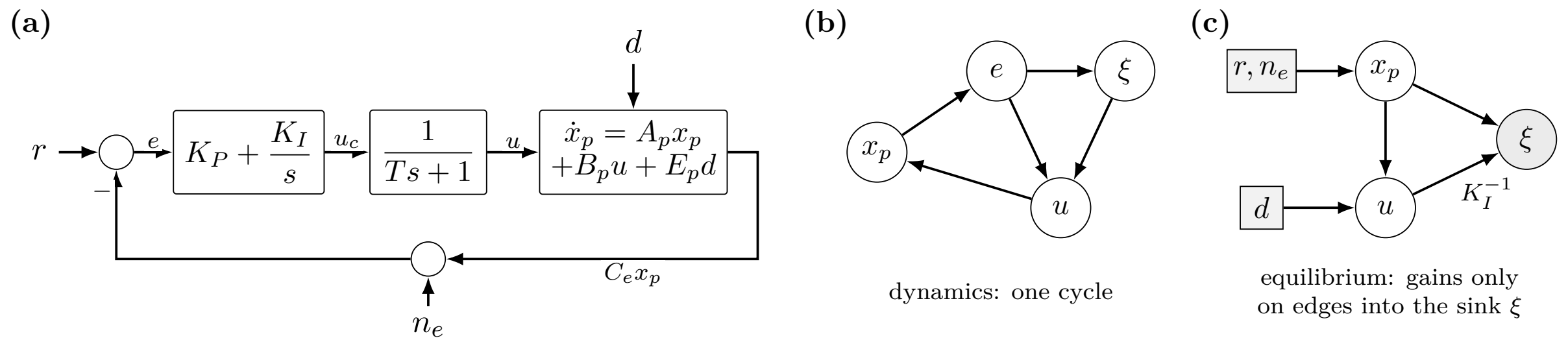

In graph terms, integral action reverses the controller's causal role at equilibrium. The regulation condition fixes the plant state, the plant fixes the input, and the controller state is left as a sink whose incoming edges are the only ones that carry gains (Figure~\ref{fig:pi}). The same holds with an internal model of any exosystem in place of the integrator: the steady state on the regulator manifold is fixed by the Francis--Wonham equations \cite{francis1977} and does not depend on the stabilising gains \cite{francis1977,singh2026c}. The discrete-time statement is identical with $\xi(k+1)=\xi(k)+e(k)$ and $I-A_d$ in place of $A$.

\subsection{Computing the equilibrium ordering}\label{sec:matching}
The equilibrium causal ordering of Section~\ref{sec:twographs} can be computed mechanically. Hold the slow variables and the noise at constant values and set every derivative to zero (in discrete time, set $x(k+1)=x(k)$). This leaves one equation for each endogenous variable. Inputs that are held fixed from outside, such as references, disturbances or a clamped integrator, are exogenous and do not appear among the unknowns.
\begin{enumerate}[label=\arabic*.]
\item \emph{Incidence graph.} Form the bipartite graph with one node per equation, one node per unknown, and an edge when the unknown appears in the equation with a nonzero coefficient. Its biadjacency matrix is the sparsity pattern of the equilibrium Jacobian.
\item \emph{Matching.} Find a perfect matching, which pairs each equation with one unknown that it is taken to determine. The Hopcroft--Karp algorithm does this in $O(E\sqrt{V})$ time for $E$ edges and $V$ nodes \cite{hopcroft1973}. If there is no perfect matching, the equilibrium is not determined for generic coefficients and the block is outside Regime~2a.
\item \emph{Orientation.} Draw an edge $x_k\to x_j$ whenever $x_k$ appears in the equation matched to $x_j$. Exogenous inputs that appear in that equation are parents of $x_j$ too.
\item \emph{Condensation.} Find the strongly connected components of the oriented graph, for example with Tarjan's algorithm in $O(V+E)$ time \cite{tarjan1972}, and sort them topologically. Each component is a set of unknowns that must be solved for together, and the resulting acyclic graph is the equilibrium causal ordering.
\end{enumerate}
A system usually has many perfect matchings, so the oriented graph in step~3 is not unique. The components and their partial order are, however: they are the irreducible blocks of the fine Dulmage--Mendelsohn decomposition, and every perfect matching produces the same condensed graph \cite{dulmage1958,murota2000}. The ordering is therefore a property of the model and not of the matching that happened to be found. The procedure uses only the sparsity pattern, so its conclusions hold for generic coefficients. A particular set of numbers can still make a block singular even when the matching exists, and this has to be checked separately (Section~\ref{sec:manystate}). The script \texttt{sinks.py} that accompanies the paper implements the four steps in about fifteen lines.

\subsection{When the controller is not a sink}\label{sec:notsink}
\begin{figure}[!htbp]
\centering
\resizebox{\textwidth}{!}{\begin{tikzpicture}[font=\small]
\begin{scope}
\node[font=\small\bfseries] at (-0.5,1.6) {(a)};
\node[ex] (r) at (0,0.9) {$r,n_e$}; \node[ex] (d) at (0,-0.5) {$d$};
\node[nd] (x) at (1.3,0.9) {$x$}; \node[nd] (u) at (1.3,-0.5) {$u$};
\node[nd,fill=black!8] (xi) at (2.6,0.2) {$\xi$};
\draw[dir] (r)--(x); \draw[dir] (x)--(u); \draw[dir] (d)--(u); \draw[dir] (x)--(xi); \draw[dir] (u)-- node[lab,below right]{\scriptsize gains}(xi);
\node[font=\scriptsize,align=center] at (1.3,-1.45) {PI: controller state\\ is the sink};
\end{scope}
\begin{scope}[xshift=4.3cm]
\node[font=\small\bfseries] at (-0.5,1.6) {(b)};
\node[ex] (r2) at (0,-0.5) {$r$}; \node[ex] (d2) at (0,0.9) {$d$};
\node[nd] (x2) at (1.5,0.9) {$x$}; \node[nd] (u2) at (1.5,-0.5) {$u$};
\draw[dir] (r2)--(u2); \draw[dir] (d2)--(x2);
\draw[dir] (x2) to[bend left=25] node[lab,right]{\scriptsize gains} (u2); \draw[dir] (u2) to[bend left=25] (x2);
\begin{scope}[on background layer]\node[scc,fit=(x2)(u2)] {};\end{scope}
\node[font=\scriptsize,align=center] at (1.0,-1.45) {P or droop: controller\\ inside the block};
\end{scope}
\begin{scope}[xshift=8.4cm]
\node[font=\small\bfseries] at (-0.5,1.6) {(c)};
\node[ex] (r3) at (0,0.9) {$r,n_e$}; \node[ex] (d3) at (0,-0.5) {$d$};
\node[nd] (x3) at (1.3,0.9) {$x$};
\node[nd] (u31) at (1.6,-0.6) {$u_1$}; \node[nd] (u32) at (2.9,-0.6) {$u_2$}; \node[nd] (xi3) at (2.9,0.6) {$\xi$};
\draw[dir] (r3)--(x3); \draw[dir] (d3)--(u31); \draw[dir] (x3)--(u31);
\draw[dir] (u31) to[bend left=15] (u32); \draw[dir] (u32) to[bend left=15] (u31);
\draw[dir] (xi3)--(u32); \draw[dir] (u32) to[bend right=30] (xi3); \draw[dir] (u31)--(xi3);
\begin{scope}[on background layer]\node[scc,fit=(u31)(u32)(xi3)] {};\end{scope}
\node[font=\scriptsize,align=center] at (1.6,-1.45) {two actuators, one integrator:\\ allocation set by gains};
\end{scope}
\begin{scope}[xshift=13.2cm]
\node[font=\small\bfseries] at (-0.5,1.6) {(d)};
\node[ex] (xs) at (0,-0.5) {$\xi_{\mathrm{sat}}$}; \node[ex] (d4) at (0,0.9) {$d,r$};
\node[nd] (x4) at (1.6,0.9) {$x$}; \node[nd] (u4) at (1.6,-0.5) {$u$};
\draw[dir] (xs)--(u4); \draw[dir] (d4)--(x4);
\draw[dir] (x4) to[bend left=25] (u4); \draw[dir] (u4) to[bend left=25] (x4);
\begin{scope}[on background layer]\node[scc,fit=(x4)(u4)] {};\end{scope}
\node[font=\scriptsize,align=center] at (1.0,-1.45) {saturated: $\xi$ becomes\\ a source};
\end{scope}
\end{tikzpicture}}
\caption{Equilibrium causal orderings of a scalar plant under four controllers; dashed boxes are strongly connected blocks, grey boxes exogenous inputs. (a) PI: $x\to u\to\xi$, the controller state is the sink. (b) Proportional or droop control: controller and plant form one block $\{u,x\}$, so the settling point depends on the gains. (c) Two actuators and one integrator: $x$ is fixed by regulation, but $\{u_1,u_2,\xi\}$ form a block, and the split of effort between actuators is set by the controller. (d) A saturated integrator is pinned at its limit and acts as a source; the ordering falls back to the proportional block.}
\label{fig:sinks}
\end{figure}
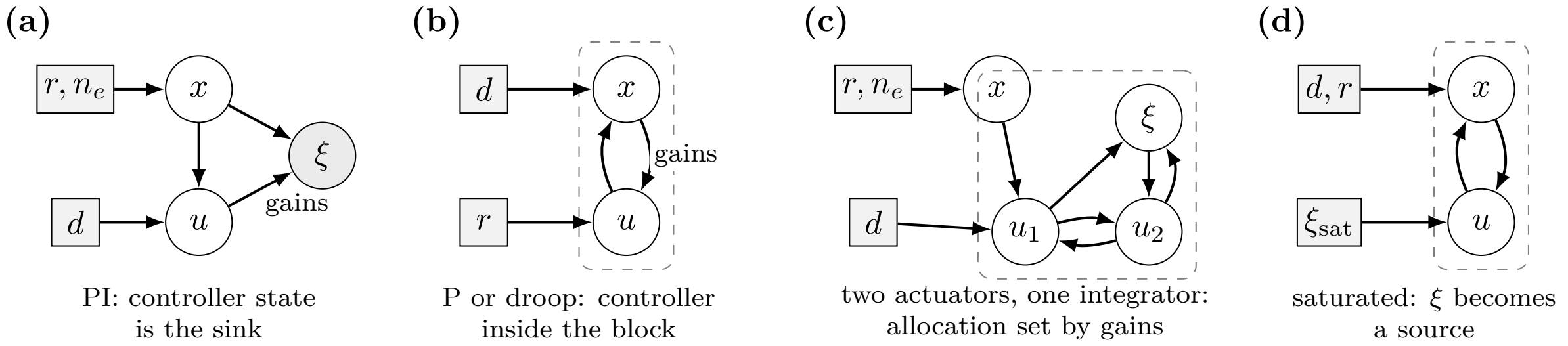

The sink property is not automatic. It holds for the equilibrium causal ordering, and only when Assumption~\ref{ass:im} holds: integral action or an internal model on the regulated errors, as many independent inputs as regulated outputs, no plant zero at $s=0$, and integrators that are not saturated. Consider the scalar plant $\dot x=-ax+bu+d$ with error $e=r-x$, $a>0$ and $b\ne0$. The four controllers below lead to four different orderings, which are drawn in Figure~\ref{fig:sinks}.
\begin{enumerate}[label=(\alph*)]
\item \emph{PI control}, $\dot\xi=e$ and $u=k_pe+k_i\xi$. The equilibrium equations are $0=r-x$, $0=-ax+bu+d$ and $u=k_p(r-x)+k_i\xi$. The matching assigns $x$ to the first, $u$ to the second and $\xi$ to the third, so the ordering is $r\to x\to u\to\xi$ with $d\to u$. The plant fixes $x^*=r$ and $u^*=(ar-d)/b$ without the gains, and $\xi^*=u^*/k_i$ is the only gain-dependent value.
\item \emph{Proportional or droop control}, $u=k_p(r-x)$. There are two equations, $0=-ax+bu+d$ and $u=k_p(r-x)$, and both contain both unknowns. Either matching closes the cycle $x\rightleftarrows u$, so $\{x,u\}$ is one block, and $x^*=(bk_pr+d)/(a+bk_p)$ depends on the gain.
\item \emph{Two actuators and one integrator}, $\dot x=-ax+b_1u_1+b_2u_2+d$ with $u_1=k_1\xi$ and $u_2=k_2\xi$. The equation $0=r-x$ is matched to $x$, as in (a). The other three equations involve $u_1$, $u_2$ and $\xi$ and form one irreducible block. Regulation holds, but the split $u_1/u_2=k_1/k_2$ comes from the controller.
\item \emph{Saturated integrator.} With $\xi$ held at its limit $\xi_{\mathrm{sat}}$, the equation $0=r-x$ is no longer enforced and $\xi_{\mathrm{sat}}$ enters from outside. What remains is $0=-ax+bu+d$ and $u=k_p(r-x)+k_i\xi_{\mathrm{sat}}$, which is the block of case (b) with one more source.
\end{enumerate}

\begin{remark}[Cases in which the controller is not a sink]\label{rem:notsink}
The main cases are these.
\begin{enumerate}[label=(\roman*)]
\item \emph{No integral action.} With proportional or droop control the controller and plant form one block at equilibrium (Figure~\ref{fig:sinks}(b)), and $J_s$ changes with the gains. In droop control this is the purpose: the droop gains are chosen to decide how load is shared.
\item \emph{More actuators than regulated outputs.} Regulation fixes the plant state and the total effort, but the allocation among actuators is set by the controller (Figure~\ref{fig:sinks}(c)). Only the regulated part of the equilibrium is gain-free.
\item \emph{Fewer actuators than regulated outputs, or a plant zero at $s=0$.} The integrators cannot all settle, or the equilibrium is not unique, and Regime~2a fails. The first shows up as a missing perfect matching; the second does not, and needs a rank test (Section~\ref{sec:manystate}).
\item \emph{Saturation.} A saturated integrator is fixed at its limit, so it becomes a source and the ordering collapses to the proportional case (Figure~\ref{fig:sinks}(d)). Anti-windup logic changes which of the two pictures applies.
\item \emph{Adaptive controllers without persistent excitation.} Parameter estimates then have a continuum of equilibria; they are not determined, and Regime~2a fails for them.
\item \emph{Outputs that read controller internals.} If a slower layer reads the integrator state itself rather than the command, it reads the one gain-dependent node (Assumption~\ref{ass:im}(e)).
\end{enumerate}
Further controller states that are driven by the regulated errors, such as filters and observers, sit downstream of the plant at equilibrium. For a PI loop with a state observer the computed ordering is $x\to u\to\hat x\to\xi$, and $\xi$ remains the sink.

Sink status is a property of the equilibrium ordering only. In the dynamic causal graph the controller always lies inside the loop's cycle, and every Regime~2b quantity, the covariance $P$, the spectrum $\Phi(\omega)$, the decay rate $\alpha$ and the inflation factor $\eta$, depends on the gains. Under Assumption~\ref{ass:im} the gains therefore decide how the system moves around its settling point, but not where that point is.
\end{remark}

\subsection{Plants with many states}\label{sec:manystate}
\begin{figure}[!htbp]
\centering
\resizebox{\textwidth}{!}{\begin{tikzpicture}[font=\small]
\begin{scope}
\node[font=\small\bfseries] at (-0.6,2.0) {(a)};
\node[ex] (r1) at (0,1.2) {$r_1$}; \node[ex] (r2) at (0,-0.2) {$r_2$};
\node[nd] (x1) at (1.2,1.2) {$x_1$}; \node[nd] (x3) at (1.2,-0.2) {$x_3$};
\node[nd] (x2) at (2.4,1.2) {$x_2$}; \node[nd] (x4) at (2.4,-0.2) {$x_4$};
\node[nd] (u1) at (3.6,1.2) {$u_1$}; \node[nd] (u2) at (3.6,-0.2) {$u_2$};
\node[nd,fill=black!8] (k1) at (4.8,1.2) {$\xi_1$}; \node[nd,fill=black!8] (k2) at (4.8,-0.2) {$\xi_2$};
\draw[dir] (r1)--(x1); \draw[dir] (r2)--(x3);
\draw[dir] (x1)--(x2); \draw[dir] (x3)--(x4); \draw[dir] (x3)--(u1);
\draw[dir] (x2)--(u1); \draw[dir] (x4)--(u2);
\draw[dir] (u1)--(k1); \draw[dir] (u2)--(k2);
\node[font=\scriptsize,align=center] at (2.4,-1.2) {sparse chain, $m=p=2$:\\ a cascade of single-variable layers};
\end{scope}
\begin{scope}[xshift=6.6cm]
\node[font=\small\bfseries] at (-0.6,2.0) {(b)};
\node[ex] (rb) at (0,0.5) {$r,d$};
\node[nd] (xb) at (1.5,1.0) {$x$}; \node[nd] (ub) at (1.5,0.0) {$u$};
\node[nd,fill=black!8] (kb1) at (3.2,1.0) {$\xi_1$}; \node[nd,fill=black!8] (kb2) at (3.2,0.0) {$\xi_2$};
\begin{scope}[on background layer]\node[scc,fit=(xb)(ub)] (B) {};\end{scope}
\draw[dir] (rb)--(B.west);
\draw[dir] (B.east|-kb1)--(kb1); \draw[dir] (B.east|-kb2)--(kb2);
\node[font=\scriptsize,align=center] at (1.6,-1.2) {dense $A,B,C$:\\ one block $\{x,u\}$, then $\xi$};
\end{scope}
\begin{scope}[xshift=11.2cm]
\node[font=\small\bfseries] at (-0.6,2.0) {(c)};
\node[ex] (rc) at (0,0.5) {$r$};
\node[nd] (xc) at (1.2,0.5) {$x$};
\node[nd] (uc2) at (2.5,-0.2) {$u_2$}; \node[nd,fill=black!8] (kc2) at (3.8,-0.2) {$\xi_2$};
\node[nd] (uc1) at (2.5,1.3) {$u_1$}; \node[nd] (uc3) at (3.6,1.3) {$u_3$}; \node[nd] (kc1) at (4.7,1.3) {$\xi_1$};
\begin{scope}[on background layer]\node[scc,fit=(uc1)(uc3)(kc1)] (Cb) {};\end{scope}
\draw[dir] (rc)--(xc); \draw[dir] (xc)--(uc2); \draw[dir] (uc2)--(kc2);
\draw[dir] (xc)--(Cb.west|-uc1); \draw[dir] (kc2) to[bend right=20] (Cb.south east);
\node[font=\scriptsize,align=center] at (2.6,-1.2) {$m=3$, $p=2$: $u_1,u_3$ act on the\\ same state; their split is set by gains};
\end{scope}
\end{tikzpicture}}
\caption{Equilibrium causal orderings for a four-state plant with $x_1$ and $x_3$ regulated; dashed boxes are strongly connected blocks, grey boxes exogenous inputs, shaded nodes sinks. (a) Sparse chain with $m=p=2$: a cascade of single-variable layers ending at $\xi_1,\xi_2$. (b) Dense $A$, $B$ and $C$: plant states and inputs form one block, and the integrators are still sinks. (c) Three inputs for two regulated outputs: $\xi_2$ is a sink, but $u_1$, $u_3$ and $\xi_1$ form a block in which the controller sets the allocation.}
\label{fig:manystate}
\end{figure}
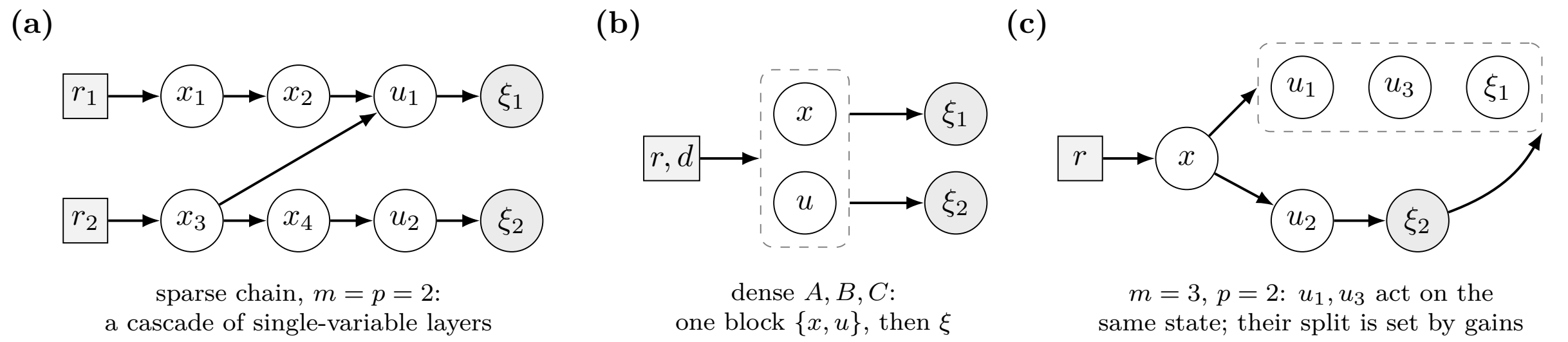

In this subsection $A$, $B$ and $C$ denote plant matrices, not those of the fast block \eqref{eq:ct}. For a plant $\dot x=Ax+Bu+Ed$ with $x\in\R^n$, $u\in\R^m$, $p$ regulated outputs $Cx=r$ and $p$ integrators with command law $u=K_I\xi$, the equilibrium equations are the $n$ plant rows, the $p$ regulation rows and the $m$ law rows, in the $n+m+p$ unknowns $(x,u,\xi)$. The procedure of Section~\ref{sec:matching} applies without change. Four things differ from the scalar case, and each is visible in the small examples of Figure~\ref{fig:manystate}, computed with the accompanying script \texttt{many\_state.py}. The plant in those examples has $n=4$ states coupled in a chain, $\dot x_i=-x_i+x_{i+1}+\dots$, with $x_1$ and $x_3$ regulated.

\emph{The plant part is not a chain.} In the scalar case $0=r-x$ fixes $x$ on its own. In general the $p$ regulation rows and the $n$ plant rows determine $(x,u)$ together, through the matrix $\left(\begin{smallmatrix}A&B\\ C&0\end{smallmatrix}\right)$ of Assumption~\ref{ass:im}(d). How finely that block splits depends on sparsity. With $u_1$ acting on $x_2$ and $u_2$ on $x_4$, every variable is its own layer: $x_1$ and $x_3$ are fixed by the references, then $x_2$ and $x_4$, then $u_1$ and $u_2$, then $\xi_1$ and $\xi_2$ (Figure~\ref{fig:manystate}(a)). Making $A$ dense merges $x_2$ and $x_4$ into one block, and making $A$, $B$ and $C$ all dense merges $x$ and $u$ into a single block of size six. In every one of these cases the integrator states come last and are sinks.

\emph{The sink is decided channel by channel.} The controller state $\xi$ is a sink in full when $m=p$ and $K_I$ is invertible. With more inputs than regulated outputs, the Dulmage--Mendelsohn decomposition of the plant part has an under-determined component, and the law rows are needed to close it. Only the redundant channels join a block with their integrators. In the example with $m=3$, where $u_1$ and $u_3$ both act on $x_2$ and $u_3=\xi_1+\xi_2$, the ordering contains $x_4\to u_2\to\xi_2$ with $\xi_2$ a sink, and a separate block $\{u_1,u_3,\xi_1\}$ in which the gains set the allocation (Figure~\ref{fig:manystate}(c)). With fewer inputs than regulated outputs ($m=1$, $p=2$) there is no perfect matching, and generic regulation is impossible. Saturation or droop in some channels affects only those channels and the blocks they are coupled to.

\emph{Structural and numerical conditions come apart.} Matching sees only which coefficients are nonzero. A transmission zero at $s=0$ makes the matrix of Assumption~\ref{ass:im}(d) singular while the sparsity pattern looks healthy. For $A=\operatorname{diag}(-1,-2)$, $B=(1,2)\T$ and $C=(1,-1)$, the matching exists and reports a block $\{x_1,x_2,u\}$ followed by $\xi$, yet $G(0)=-CA^{-1}B=0$ and the Rosenbrock determinant \cite{rosenbrock1970state} vanishes, so the equilibrium is not unique. With many states the structural check therefore has to be paired with a rank test on $\left(\begin{smallmatrix}A&B\\ C&0\end{smallmatrix}\right)$. In the scalar case that test reduces to $b\ne0$.

\emph{Size is not an obstacle.} Matching and condensation are close to linear in the number of nonzero coefficients, so plants with thousands of states are routine. The layered ordering that results, for instance a DC link that sets the current references of inner loops which in turn set the integrators of an outer controller, is the equilibrium counterpart of the slow--fast cascade that condensation exploits (Section~\ref{sec:separation}).

\section{Regime 2b: induced confounding}\label{sec:2b}

\subsection{The stationary covariance}
Let the block be in Regime~2b (Definition~\ref{def:2b}). With $x_s$ held fixed, the fluctuation of the block about its equilibrium has a unique stationary distribution with covariance $P$ solving
\begin{equation}\label{eq:lyap}
AP+PA\T+BB\T=0\qquad\text{(continuous-time Lyapunov equation)},
\end{equation}
\begin{equation}\label{eq:stein}
P=A_dPA_d\T+Q_d\qquad\text{(discrete-time Lyapunov, or Stein, equation)}.
\end{equation}
The solutions are $P=\int_0^\infty e^{At}BB\T e^{A\T t}\,dt$ and $P=\sum_{k\ge0}A_d^kQ_d(A_d\T)^k$. For Gaussian noise the stationary distribution is Gaussian with this covariance.

\begin{definition}[Induced confounding]\label{def:kappa}
In Regime~2b the \emph{induced confounding} between outputs $i$ and $j$ is $\kappa_{ij}=c_i\T Pc_j$, where $c_i\T$ is the $i$th row of $C$. Its frequency-resolved form is the cross-spectrum
\[
\Phi(\omega)=T(j\omega)T(j\omega)^{*},\qquad T(s)=C(sI-A)^{-1}B+D_w,
\]
and in discrete time $\Phi(\theta)=T(e^{j\theta})T(e^{j\theta})^{*}$ with $T(z)=C(zI-A_d)^{-1}G$, where $Q_d=GG\T$.
\end{definition}

Marginalising the fast block therefore leaves two kinds of edge (Figure~\ref{fig:scheme}(c)). Directed edges $x_s\to y_i$ carry $J_s$, which is Regime~2a content. Bidirected edges $y_i\leftrightarrow y_j$ carry $\kappa_{ij}$ or $\Phi_{ij}(\omega)$, which is Regime~2b content. The same covariance $P$ appears in graphical continuous Lyapunov models \cite{varando2020,dettling2023}, where it is the object to be fitted. Here it is the object to be bounded and designed.

\subsection{Sampling maps one equation onto the other}
\begin{proposition}\label{prop:sampling}
Let $x_f$ obey $\dot x_f=Ax_f+B\dot w$ with $A$ Hurwitz, and let $x(k)=x_f(kh)$. Then $x(k+1)=A_dx(k)+w(k)$ with $A_d=e^{Ah}$ and $Q_d=\int_0^he^{As}BB\T e^{A\T s}ds$; $A_d$ is Schur; and the solution of the Stein equation \eqref{eq:stein} equals the solution of the Lyapunov equation \eqref{eq:lyap}.
\end{proposition}

\begin{proof}
The first two claims are the standard exact discretisation of a linear stochastic differential equation \cite{astrom1970}, and $\rho(e^{Ah})=e^{h\alpha}<1$. For the third, $e^{Ah}Pe^{A\T h}+Q_d=\int_h^\infty e^{At}BB\T e^{A\T t}dt+\int_0^he^{At}BB\T e^{A\T t}dt=P$, so $P$ solves \eqref{eq:stein}, whose solution is unique when $A_d$ is Schur.
\end{proof}

A sampled system therefore inherits its induced confounding exactly from the continuous one. Numerically, for a three-stage chain the two solutions agree to within $3\times10^{-14}$ in relative terms at $h=0.1$, $0.01$ and $0.001$ (Example~\ref{ex:chain}). Sampling also limits how a block can leave Regime~2b. A real eigenvalue $\lambda$ of $A$ maps to $e^{\lambda h}>0$, and a negative real eigenvalue of $e^{Ah}$ can come only from a complex pair of $A$ with $\operatorname{Im}\lambda\,h$ an odd multiple of $\pi$, in which case it is double. A simple eigenvalue crossing $-1$, the flip, therefore cannot arise from sampling. Discrete-time systems that are not sampled continuous ones, such as over-tuned digital controllers, can have one, and this gives them an exit from Regime~2b that continuous systems do not have (Section~\ref{sec:exits}).

\subsection{Leaving Regime 2b while staying in Regime 2a}\label{sec:exits}
\begin{proposition}\label{prop:exits}
Let a family of blocks depend continuously on a parameter and leave Regime~2b at a parameter value where it is still in Regime~2a. Then at that value an eigenvalue of $A$ lies on the imaginary axis away from the origin (continuous time), or an eigenvalue of $A_d$ lies on the unit circle away from $1$ (discrete time). The condensed sensitivity $J_s$ stays finite and continuous through the crossing, while $\|P\|\to\infty$ as the crossing is approached from inside Regime~2b, provided the critical mode is excited by the noise.
\end{proposition}

\begin{proof}
Leaving the Hurwitz (Schur) set means that an eigenvalue reaches the boundary of that set, and Regime~2a rules out the boundary point $0$ (respectively $1$). $J_s$ depends on $A^{-1}$ (respectively $(I-A_d)^{-1}$), which is continuous where it exists. For the divergence, take a left eigenvector $v$ of the critical eigenvalue $\lambda$, normalised. Then
\[
v^*Pv=\int_0^\infty e^{2\operatorname{Re}\lambda t}\,v^*BB\T v\,dt=\frac{v^*BB\T v}{2|\operatorname{Re}\lambda|}
\]
in continuous time and $v^*Pv=v^*Q_dv/(1-|\lambda|^2)$ in discrete time, both unbounded when $v^*BB\T v>0$ (respectively $v^*Q_dv>0$).
\end{proof}

\begin{figure}[!htbp]
\centering
\begin{tikzpicture}[font=\small]
\begin{scope}
\node[font=\small\bfseries] at (-2.4,2.2) {(a)};
\fill[black!10] (-2.2,-1.9) rectangle (0,1.9);
\draw[-{Latex[length=2mm]}] (-2.3,0) -- (2.0,0) node[right]{$\operatorname{Re}\lambda$};
\draw[-{Latex[length=2mm]}] (0,-2.0) -- (0,2.1) node[above]{$\operatorname{Im}\lambda$};
\draw[thick] (0,-1.9) -- (0,1.9);
\fill[white] (0,0) circle (3pt); \draw[thick] (0,0) circle (3pt);
\node[align=center,font=\scriptsize] at (-1.15,1.35) {Regime 2b\\ (Hurwitz)};
\node[align=left,font=\scriptsize] at (1.05,1.35) {2a only:\\ $\det A\neq0$};
\draw[-{Latex[length=1.6mm]}] (-0.7,0.9) to[bend left=20] node[pos=0.95,right,font=\scriptsize]{Hopf} (-0.05,1.0);
\draw[-{Latex[length=1.6mm]}] (-0.8,-0.35) -- node[below,font=\scriptsize]{fold} (-0.12,-0.05);
\node[font=\scriptsize,align=center] at (0.2,-1.5) {origin excluded:\\ 2a lost};
\end{scope}
\begin{scope}[xshift=6.6cm]
\node[font=\small\bfseries] at (-2.6,2.2) {(b)};
\fill[black!10] (0,0) circle (1.5);
\draw[-{Latex[length=2mm]}] (-2.3,0) -- (2.3,0) node[right]{$\operatorname{Re}\lambda$};
\draw[-{Latex[length=2mm]}] (0,-2.0) -- (0,2.1) node[above]{$\operatorname{Im}\lambda$};
\draw[thick] (0,0) circle (1.5);
\fill[white] (1.5,0) circle (3pt); \draw[thick] (1.5,0) circle (3pt);
\node[font=\scriptsize] at (1.75,-0.3) {$1$};
\node[align=center,font=\scriptsize] at (-0.45,0.55) {Regime 2b\\ ($\rho<1$)};
\draw[-{Latex[length=1.6mm]}] (-0.9,-0.25) -- node[below,font=\scriptsize]{flip} (-1.42,-0.05);
\draw[-{Latex[length=1.6mm]}] (0.6,0.9) -- node[right,font=\scriptsize,pos=0.9]{\,NS} (0.95,1.13);
\node[align=left,font=\scriptsize] at (2.0,1.6) {2a only:\\ $\det(I-A_d)\neq0$};
\end{scope}
\end{tikzpicture}
\caption{Regime~2b (shaded) and Regime~2a in the eigenvalue plane. (a) Continuous time: Regime~2a excludes only the origin; Regime~2b is the open left half-plane. A Hopf crossing leaves 2b and keeps 2a; a fold leaves both. (b) Discrete time: Regime~2a excludes only $\lambda=1$; Regime~2b is the open unit disk. Flip and Neimark--Sacker (NS) crossings leave 2b and keep 2a.}
\label{fig:regimes}
\end{figure}
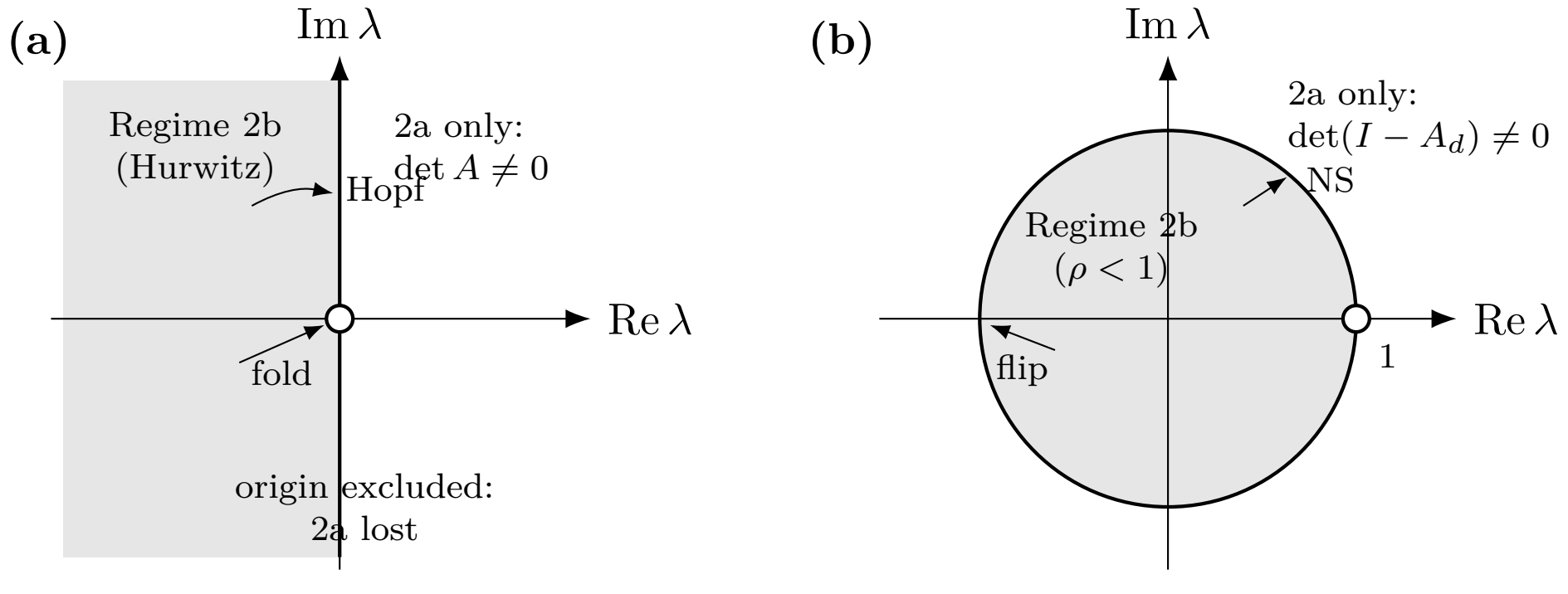

In continuous time the generic exit with the equilibrium intact is a Hopf crossing. In discrete time there are two: a flip, where a real eigenvalue passes $-1$, and a Neimark--Sacker crossing, where a complex pair crosses the unit circle. Figure~\ref{fig:regimes} shows the regions, and Examples~\ref{ex:hopf} and~\ref{ex:discrete} work each exit. A fold, where a real eigenvalue reaches $0$ or $1$, leaves both regimes at once. Not every bidirected edge need diverge at a Hopf exit: a pair of outputs in quadrature with the critical oscillation can keep a bounded instantaneous covariance while their cross-spectrum at the critical frequency diverges (Example~\ref{ex:hopf}). This is one reason to keep the frequency-resolved edges.

\subsection{The confounding gain}
How large can the induced edges be? The answer depends on two things that are often confused: how slowly the block forgets, and how strongly its non-normal couplings amplify disturbances on the way.

\begin{definition}[Confounding gain]\label{def:gain}
In energy coordinates, the \emph{unit-noise Gramian} is the solution $\Pi$ of $\tilde A\Pi+\Pi\tilde A\T+I=0$ in continuous time, and of $\Pi_d=\tilde A_d\Pi_d\tilde A_d\T+I$ in discrete time. The \emph{confounding gain} is $\|\Pi\|$ (respectively $\|\Pi_d\|$). With $\alpha=\max_i\operatorname{Re}\lambda_i(A)$ and $\rho=\rho(A_d)$, define the \emph{inflation factors}
\[
\eta=2|\alpha|\,\|\Pi\|,\qquad \eta_d=(1-\rho^2)\|\Pi_d\| ,
\]
so that
\begin{equation}\label{eq:gain}
\|\Pi\|=\frac{\eta}{2|\alpha|},\qquad \|\Pi_d\|=\frac{\eta_d}{1-\rho^2} .
\end{equation}
\end{definition}

\begin{lemma}\label{lem:factor}
$\eta\ge1$ and $\eta_d\ge1$, with equality when $\tilde A$ (respectively $\tilde A_d$) is normal.
\end{lemma}

\begin{proof}
Let $v$ be a unit left eigenvector of $\tilde A$ for an eigenvalue with real part $\alpha$. Then $v^*\tilde A=\lambda v^*$ and $v^*\Pi v=\int_0^\infty|e^{\lambda t}|^2dt=1/(2|\alpha|)$, so $\|\Pi\|\ge1/(2|\alpha|)$. In discrete time the same argument with $v^*\tilde A_d^k=\lambda^kv^*$ gives $v^*\Pi_dv=1/(1-|\lambda|^2)$. For normal $\tilde A$, $\Pi$ is diagonal in the eigenbasis with entries $1/(2|\operatorname{Re}\lambda_i|)$, whose largest is $1/(2|\alpha|)$; the discrete case is the same.
\end{proof}

The factor $1/(2|\alpha|)$, or $1/(1-\rho^2)$, is critical slowing down: fluctuations linger when the least-damped mode is near the boundary. It is the same mechanism behind early-warning signals of critical transitions \cite{scheffer2009}. The factor $\eta$ is non-normal amplification: skewed eigenvectors let disturbances grow before they decay \cite{trefethen2005}. A design change can move the two factors in opposite directions, and only their product governs the confounding.

\begin{theorem}[Worst-case amplification]\label{thm:gain}
In Regime~2b, every solution $X$ of $\tilde AX+X\tilde A\T+E=0$ with $E$ symmetric satisfies $\|X\|\le\|\Pi\|\,\|E\|$, with equality for $E=I$. The same holds for $X=\tilde A_dX\tilde A_d\T+E$ with $\Pi_d$ in place of $\Pi$.
\end{theorem}

\begin{proof}
The maps $E\mapsto X=\int_0^\infty e^{\tilde At}Ee^{\tilde A\T t}dt$ and $E\mapsto X=\sum_kA^k_dE(A_d\T)^k$ (in energy coordinates) are linear and send positive semidefinite matrices to positive semidefinite matrices. From $-\|E\|I\preceq E\preceq\|E\|I$ it follows that $-\|E\|\Pi\preceq X\preceq\|E\|\Pi$, hence $\|X\|\le\|E\|\|\Pi\|$; the discrete case is identical. Equality at $E=I$ is the definition of $\Pi$ (Definition~\ref{def:gain}).
\end{proof}

For $2000$ random stable $4\times4$ matrices $A_d$ and random symmetric $E$, the ratio $\|\Pi_d\|\|E\|/\|X\|$ never fell below $1.035$, and $\eta_d\ge1$ held in every case. The covariance of outputs and its bound follow at once.

\begin{corollary}[Edge bound]\label{cor:edge}
For outputs $y_i=c_i\T x_f$, $|\kappa_{ij}|\le\|Q^{-1/2}c_i\|\,\|Q^{-1/2}c_j\|\,\|\Pi\|\,\|\tilde B\tilde B\T\|$, and likewise in discrete time with $\Pi_d$ and $\tilde Q_d$.
\end{corollary}

\begin{proof}
Apply Theorem~\ref{thm:gain} to $\tilde P=Q^{1/2}PQ^{1/2}$ with $E=\tilde B\tilde B\T$, and use
\[
\kappa_{ij}=(Q^{-1/2}c_i)\T\tilde P(Q^{-1/2}c_j)
\]
together with the Cauchy--Schwarz inequality.
\end{proof}

The confounding gain depends on the storage metric, while $\kappa$ does not. Statements that use $\|\Pi\|$ should therefore name the metric and be checked against other admissible metrics; a design rule for that is given in \cite{singh2026c}. A sampled system has $\Pi_d=\sum_ke^{\tilde Akh}e^{\tilde A\T kh}$, a Riemann sum of $\Pi/h$, so $h\|\Pi_d\|\to\|\Pi\|$ as $h\to0$; for the chain of Example~\ref{ex:chain}, $h\|\Pi_d\|$ is $5.639$, $5.594$ and $5.589$ at $h=0.1$, $0.01$ and $0.001$, against $\|\Pi\|=5.589$.

\subsection{The fluctuation--dissipation defect}
When is there no induced confounding to bound? In energy coordinates the answer is a balance between the noise a block receives and the dissipation it provides.

\begin{proposition}[Defect]\label{prop:defect}
In continuous time let $R=-(\tilde A+\tilde A\T)/2$, and for $k\ge0$ write $\tilde P=kI+\Delta_k$. Then $\tilde A\Delta_k+\Delta_k\tilde A\T+E_k=0$ with $E_k=\tilde B\tilde B\T-2kR$. In discrete time, with $\tilde P=kI+\Delta_k$, $\Delta_k=\tilde A_d\Delta_k\tilde A_d\T+E_k$ with $E_k=\tilde Q_d-k(I-\tilde A_d\tilde A_d\T)$. If the noise is nonzero and $R$ is not positive semidefinite (respectively $\|\tilde A_d\|>1$), then $E_k\ne0$ for every $k\ge0$.
\end{proposition}

\begin{proof}
Substitute $\tilde P=kI+\Delta_k$ into \eqref{eq:lyap} or \eqref{eq:stein}, written in energy coordinates. For $k=0$, $E_0$ is the noise covariance, which is nonzero. For $k>0$, $E_k=0$ would make $\tilde B\tilde B\T=2kR$, or $\tilde Q_d=k(I-\tilde A_d\tilde A_d\T)$; the left side is positive semidefinite and the right side is not.
\end{proof}

The defect $E_k$ measures how far the block is from fluctuation--dissipation balance \cite{kubo1966}, in which noise is injected exactly where and as much as energy is dissipated. A block that can increase stored energy transiently, $R\not\succeq0$ in continuous time or $\|\tilde A_d\|>1$ in one discrete step, can never be in balance, and Theorem~\ref{thm:gain} then says that $\|\Pi\|$ is the worst-case factor by which its defect becomes confounding. Control delay is a common cause of such transient growth; in a grid-following converter it makes the fast loops non-passive in any separable storage \cite{singh2026a}.

\subsection{Interventions}\label{sec:interv}
Two kinds of intervention behave differently.

\emph{Intervening on a slow variable}, $\mathrm{do}(x_s=v)$, leaves the fast block unchanged. Its equilibrium moves along the condensed map, $y\mapsto J_sv$, and its stationary fluctuation, and hence every bidirected edge, is unaffected, because $P$ does not depend on $x_s$. In Regime~2b the dynamic and the condensed descriptions agree on such interventions: the block relaxes to the new equilibrium at a rate set by $\alpha$, or by $\rho$ in discrete time.

\emph{Intervening on a fast variable}, clamping a state of the block, removes that state's equation and turns it into an input. The intervened block has a different matrix $A_{\setminus i}$ and may fall in a different regime: cutting a loop can stabilise a block or destabilise it. The equilibrium of the intervened dynamics coincides with the intervened equilibrium model when both are in Regime~2a and the intervened dynamics converge \cite{mooij2013}; this is where the caveat of \cite{dash2005} becomes concrete. Regime~2b of the intervened block, needed for its stationary edges, must be checked separately.

\subsection{Identification from data}
Bidirected edges make ambient data misleading about directed effects. The following is the closed-loop form of a familiar statement \cite{forssell1999}.

\begin{proposition}[Least squares and instruments]\label{prop:id}
Let $y=G(s)u+v$, where the disturbance $v$ is correlated with $u$ through feedback. The least-squares estimate $\Phi_{yu}\Phi_{uu}^{-1}$ equals $G+\Phi_{vu}\Phi_{uu}^{-1}$. If an excitation $r$ uncorrelated with $v$ is injected and $\Phi_{ur}$ is invertible, the instrumental-variable estimate $\Phi_{yr}\Phi_{ur}^{-1}$ equals $G$.
\end{proposition}

\begin{proof}
From $y=Gu+v$, $\Phi_{yu}=G\Phi_{uu}+\Phi_{vu}$ and $\Phi_{yr}=G\Phi_{ur}+\Phi_{vr}=G\Phi_{ur}$.
\end{proof}

The bias term $\Phi_{vu}\Phi_{uu}^{-1}$ is the induced confounding seen from the estimator's side. Example~\ref{ex:regression} shows that it can reverse the sign of a causal effect. For a slower layer that estimates a quantity from several fast-block outputs, the same edges set how much information those outputs carry: the Fisher information must be computed from sensor noise plus the induced spectrum, and using sensor noise alone overstates it, by a factor of about ten in a converter and about a hundred in a vehicle platoon \cite{singh2026c}.

\section{Condensation at finite time-scale separation}\label{sec:separation}

So far $x_s$ has been held fixed. When a slower layer updates it, $\tau\dot x_s=-k\T(y-y^\ast)$ with $k\T J_s=1$, condensation is an approximation whose accuracy depends on $\tau$. Let $L(s)=C(sI-A)^{-1}H+D_p$, so that $L(0)=J_s$. The results below are proved in \cite{singh2026c}; we state them for reference.

\begin{proposition}[Certificate]\label{prop:cert}
Let $\Delta_r=\{\lambda:|\lambda+1/\tau|\le r/\tau\}$, $0<r\le1$, contain no eigenvalue of $A$. If $\max_{\partial\Delta_r}|k\T L(\lambda)-1|<r$, the full system has exactly one eigenvalue in $\Delta_r$.
\end{proposition}

\begin{proposition}[A-priori bound]\label{prop:apriori}
Let $M=\sup_{t\ge0}\|e^{\tilde At}\|e^{|\alpha|t/2}$ and $\rho_c=\|k\T\tilde C\|\|\tilde A^{-1}\tilde H\|$. If $\tau|\alpha|>4$ and $\delta=4\rho_cM/(\tau|\alpha|-4)<1$, the full system has exactly one eigenvalue within $\delta/\tau$ of $-1/\tau$.
\end{proposition}

\begin{proposition}[First order]\label{prop:first}
In Regime~2b, with $\beta=-k\T CA^{-2}H$, $\Sigma(\omega)$ the spectrum of $y$ at fixed $x_s$, and
\[
\nu=\frac{2}{\pi\,k\T\Sigma(0)k}\int_0^\infty\frac{k\T(\Sigma(\omega)-\Sigma(0))k}{\omega^2}\,d\omega,
\]
the slow eigenvalue is $-1/(\tau+\beta)+O(\tau^{-3})$ and $\operatorname{Var}x_s=\frac{k\T\Sigma(0)k}{2\tau}\bigl(1+\frac{\nu-\beta}{\tau}\bigr)+o(\tau^{-2})$. A channel delay $T$ replaces $\beta$ by $\beta-T$.
\end{proposition}
\begin{corollary}[Separation needed for the noise]\label{cor:noise-separation}
Under the conditions of Proposition~\ref{prop:first}, let a channel delay $T\ge 0$
act on the slow layer, and let $\varepsilon_{\mathrm{tol}}\in(0,1)$ be the admissible
relative error in $\operatorname{Var} x_s$. To first order in $1/\tau$, the condensed
prediction $k^\top\Sigma(0)k/(2\tau)$ meets this tolerance if
\begin{equation}\label{eq:noise-separation}
  \tau \;\ge\; \frac{\lvert \nu-\beta+T\rvert}{\varepsilon_{\mathrm{tol}}}.
\end{equation}
The slow eigenvalue meets the same tolerance if
$\tau \ge \lvert \beta - T\rvert/\varepsilon_{\mathrm{tol}}$.
\end{corollary}

\begin{proof}
By Proposition~\ref{prop:first}, with $\beta$ replaced by $\beta-T$, the relative
error of the condensed variance is $(\nu-\beta+T)/\tau + o(\tau^{-1})$, and that of the
slow eigenvalue is $(\beta-T)/\tau + O(\tau^{-2})$. Bounding each by
$\varepsilon_{\mathrm{tol}}$ gives the two conditions.
\end{proof}
The eigenvalue condenses with error $\beta/\tau$, the noise with error $(\nu-\beta)/\tau$. When the induced spectrum rises quickly away from zero frequency, the noise needs much more separation than the dynamics. In a vehicle platoon the slow eigenvalue was accurate to 1\% at $\tau|\alpha|=5$, while the slow layer's variance needed $\tau|\alpha|\approx76$ for 10\% accuracy; in a converter, $8$ to $14$ sufficed \cite{singh2026c}. This is the dynamic content of Regime~2b that equilibrium models do not have.

\section{Regime 2b and normally attracting, asymptotically invariant manifolds}\label{sec:naim}

\subsection{What a NAIM is}
A Normally Attracting, Invariant Manifold (NAIM) is a surface in the state space with two properties. The system stays on it once it gets there, which is invariance; in the asymptotic version the surface is only approached as time runs on, or as the separation of time scales grows. And states slightly off the surface are pulled back towards it in the directions across it, which is normal attraction. Motion along the surface, the tangential or internal dynamics, is a separate matter: it can be stable, exponentially stable or unstable without affecting normal attraction.

The heated room of Section~\ref{sec:orientation} gives a picture. For each outside temperature there is one settled pair of room temperature and valve opening, and these pairs form a curve in the state space of the loop. Disturbances knock the state off the curve and the loop pulls it back; that is normal attraction. Slow changes in the outside temperature move the state along the curve; that is the tangential dynamics. The NAIM research programme uses this picture as a common language for slow--fast systems, accelerated optimisation and control \cite{naim2026,fenichel1979}.

\subsection{How the regimes sit inside it}
For the fast block \eqref{eq:ct} the settled response $x_f=h(x_s)$, linear here, $h(x_s)=-A^{-1}Hx_s$, is the candidate manifold, and the two regimes are statements about it.

\emph{Regime 2a is existence and uniqueness of the manifold.} The settled response is a well-defined graph over the slow variables exactly when the normal linearisation has no zero eigenvalue, which is the implicit-function condition, $\det A\ne0$ (or $\det(I-A_d)\ne0$). This is what cyclic structural causal models need in order to marginalise the block.

\emph{Regime 2b is normal attraction of that manifold.} All normal eigenvalues lie in the open left half-plane (or inside the unit disk). The exits of Section~\ref{sec:exits} are the places where the manifold survives as a set of equilibria but stops attracting: a Hopf, flip or Neimark--Sacker crossing turns some normal directions from attracting to repelling, while the graph $x_f=h(x_s)$, and with it the condensed causal graph, is unchanged.

\emph{Asymptotic invariance is condensation at finite separation.} When the slow layer moves, $x_s$ is not constant, and the manifold is invariant only in the limit of infinite separation. Fenichel's theory guarantees that a normally attracting manifold persists, displaced by an amount of order $\varepsilon$, when the slow variables move at rate $\varepsilon$ \cite{fenichel1979}. 

Proposition~\ref{prop:first} is the quantitative linear version, with $\varepsilon=1/(\tau|\alpha|)$: the slow dynamics see the manifold shifted by the lag $\beta$, and the slow layer's noise by $\nu-\beta$.

\emph{Noise is what Regime 2b adds to the geometry.} A NAIM says that the state returns to the manifold. With noise it never quite does: it wanders in a stochastic tube around the manifold, and the Lyapunov equation \eqref{eq:lyap}, or the Stein equation \eqref{eq:stein}, gives the shape of that tube in the normal directions. The confounding gain $\|\Pi\|=\eta/(2|\alpha|)$ measures how strong the normal attraction really is once transient growth is counted. The factor $1/(2|\alpha|)$ is the slowest rate of return. The factor $\eta$ is the extra spread caused by non-normal coupling between normal directions, which in the geometric language of the NAIM programme is a tilt of the fibres along which states return. The bidirected edges are the causal reading of the tube: outputs that both see a normal direction wander together.

\emph{Internal models fix the manifold, gains fix the attraction.} Proposition~\ref{prop:gainfree}, and its Francis--Wonham form \cite{francis1977}, say that with integral action or an internal model the location of the manifold, the regulator manifold, does not depend on the controller gains. The gains decide only how strongly and how cleanly the manifold attracts. In NAIM terms, the design of a regulating loop separates into placing the manifold and shaping its normal attraction.

So Regime~2b does follow from the NAIM picture in a definite sense. It is normal attraction of the settled response, together with the stationary statistics of the fluctuations about it and their causal reading. In the linear setting of this paper the correspondence is exact. In the nonlinear setting it holds locally, with the qualifications of Section~\ref{sec:scope}.

\section{What attraction adds: eight points}\label{sec:points}

The step from Regime~2a to Regime~2b is easy to describe loosely and easy to describe wrongly. This section collects the distinctions in one place. Each point gives a statement and then what it implies for analysis or design.

The dividing line is attraction, not hyperbolicity. A hyperbolic block has no eigenvalue on the imaginary axis (no eigenvalue on the unit circle in discrete time), but it may have eigenvalues in the right half-plane. A saddle is hyperbolic, and it is still never observed: trajectories leave it, and its Lyapunov equation~\eqref{eq:lyap} has a unique solution, but when the noise reaches the unstable modes that solution is not positive semidefinite and so is not a covariance. Its slow manifold persists under perturbation in Fenichel's sense, but nothing stays on it. Regime~2b requires $A$ Hurwitz ($A_d$ Schur), which is normal attraction in the language of Section~\ref{sec:naim}. In what follows, ``2a without 2b'' means a block that is uniquely solvable but not attracting. Each point is illustrated by a figure. Unless stated otherwise, the figures use the two-state loop \eqref{eq:two1} with $a=-2$ and $b=1$, which is in Regime~2a for every damping $d$ and in Regime~2b for $d>0$; all numbers in them were computed with the accompanying script \texttt{points.py}.

\begin{enumerate}[label=\arabic*.,leftmargin=*]
\item \emph{The equilibrium exists but is not where the system settles.} A block in Regime~2a without 2b has a unique equilibrium, and the block does not converge to it. At an exact Hopf point of a linear block, the time average of the oscillation still equals the equilibrium, so the equilibrium is seen only on average.

\emph{Implication.} Long-run averages of a marginal block can look correct while every snapshot is far from the predicted value. Agreement of averages with the equilibrium model is not evidence of Regime~2b.

\ptfig{\input{tikz/pt1.tex}}{Point 1 on the two-state loop with $a=-2$, $b=1$ and a unit step in $u$ entering $x_1$. (a) The dynamic graph is a single cycle, while the equilibrium model $u\to x^*$ is defined for every $d$. (b) Response of $x_2$. Black: $d=0.3$, settling to $x_2^*=0.478$. Grey: $d=0$, a Hopf point, oscillating between $0$ and $1$ and never settling. Dashed: running time average for $d=0$, which tends to its equilibrium value $0.5$ (dotted).}{fig:pt1}

\item \emph{The condensed model still predicts the shift of the equilibrium.} For interventions that leave the loop intact, such as $\mathrm{do}(x_s=v)$, the condensed sensitivity $J_s$ remains finite and continuous through a Hopf, flip or Neimark--Sacker crossing (Proposition~\ref{prop:exits}). What fails is the prediction that the system goes to the shifted equilibrium; the dynamic and condensed descriptions agree on that only inside Regime~2b (Section~\ref{sec:interv}). Interventions that cut the loop are no safer: clamping a fast variable changes $A$ to $A_{\setminus i}$, which can stabilise or destabilise the block, so Regime~2b has to be checked again for the intervened block.

\emph{Implication.} An equilibrium model is not wrong about where the equilibrium moves. It is silent about whether the system follows, and that question belongs to Regime~2b.

\ptfig{\begin{tikzpicture}[font=\small]
\begin{scope}
\node[font=\small\bfseries] at (-0.7,1.4) {(a)};
\node[ex] (v) at (0,0.5) {$\mathrm{do}(u=v)$};
\node[nd] (y) at (2.6,0.5) {$x_2^*$};
\draw[dir] (v)-- node[lab,above]{\scriptsize $J_s$} (y);
\node[font=\scriptsize,align=center] at (1.3,-0.6) {loop intact: the shift $J_s\,\Delta v$ is right\\ in Regime~2a; reaching it needs 2b};
\end{scope}
\begin{scope}[xshift=5.6cm]
\node[font=\small\bfseries] at (-0.7,1.4) {(b)};
\node[nd] (a1) at (0.3,1.0) {$x_1$}; \node[nd] (a2) at (0.3,-0.2) {$x_2$};
\draw[dir] (a1) to[bend left=30] (a2); \draw[dir] (a2) to[bend left=30] (a1);
\begin{scope}[on background layer]\node[scc,fit=(a1)(a2)] {};\end{scope}
\draw[-{Latex[length=2.5mm]},thick] (1.1,0.4)--(2.1,0.4);
\node[ex] (c1) at (3.0,1.0) {$x_1{=}v$}; \node[nd] (c2) at (3.0,-0.2) {$x_2$};
\draw[dir] (c1)-- node[lab,right]{\scriptsize $b$} (c2);
\draw[dir] (c2) edge[loop right,looseness=6] node[lab,right]{\scriptsize $-d$} (c2);
\node[font=\scriptsize,align=center] at (1.7,-1.25) {clamping $x_1$ cuts the loop: $\dot x_2=bv-dx_2$;\\ $d>0$: Regime~2b; $d=0$: an integrator, outside 2a};
\end{scope}
\end{tikzpicture}}{Point 2. (a) For an intervention that leaves the loop intact, the condensed edge $J_s$ gives the correct shift of the equilibrium in Regime~2a; whether the system gets there is a Regime~2b question. (b) Clamping $x_1$ cuts the loop and leaves $\dot x_2=bv-dx_2$, a new block whose regime must be checked again: for $d=0$ the intervened block is a pure integrator and is in neither regime.}{fig:pt2}

\item \emph{Condensation is defined; its use by a slower layer is not justified.} Marginalising a cycle in the sense of \cite{bongers2021} needs only unique solvability, and that survives in Regime~2a. What fails is the justification for a slower layer to treat the block as condensed. The certificate and the a-priori bound of Section~\ref{sec:separation} need $|\alpha|>0$, and the first-order corrections scale with $1/(\tau|\alpha|)$.

\emph{Implication.} The condensed graph of a non-attracting block can still be drawn and reasoned about at equilibrium, but a supervisory layer must not be designed on it.

\ptfig{\begin{tikzpicture}[font=\small]
\begin{scope}
\node[font=\small\bfseries] at (-0.8,1.6) {(a)};
\node[nd] (s) at (0,0.4) {$x_s$};
\node[nd] (f1) at (1.7,1.0) {$x_1$}; \node[nd] (f2) at (1.7,-0.2) {$x_2$};
\draw[dir] (s)--(f1);
\draw[dir] (f1) to[bend left=30] (f2); \draw[dir] (f2) to[bend left=30] (f1);
\begin{scope}[on background layer]\node[scc,fit=(f1)(f2)] (F) {};\end{scope}
\draw[dir] (f2) to[bend left=35] node[lab,below]{\scriptsize $-k/\tau$} (s);
\node[font=\scriptsize,align=center] at (0.9,-1.25) {full model: slow integrator\\ over the fast loop};
\end{scope}
\begin{scope}[xshift=4.3cm]
\node[font=\small\bfseries] at (-0.8,1.6) {(b)};
\node[nd] (s2) at (0,0.4) {$x_s$}; \node[nd] (y2) at (1.8,0.4) {$y$};
\draw[dir] (s2) to[bend left=30] node[lab,above]{\scriptsize $J_s$} (y2);
\draw[dir] (y2) to[bend left=30] node[lab,below]{\scriptsize $-k/\tau$} (s2);
\node[font=\scriptsize,align=center] at (0.9,-1.25) {condensed model:\\ slow eigenvalue $-1/\tau$};
\end{scope}
\begin{scope}[xshift=8.6cm]
\node[font=\small\bfseries] at (-0.5,1.6) {(c)};
\node[font=\scriptsize,anchor=north west,align=left] at (-0.3,1.3) {
\begin{tabular}{@{}lcc@{}}
\toprule
 & $d=0.3$ & $d=0$\\
\midrule
slow eigenvalue & $-0.1025$ & $-0.0995$\\
fast pair, real part & $-0.249$ & $+0.050$\\
condensed prediction & $-0.1$ & $-0.1$\\
\bottomrule
\end{tabular}};
\node[font=\scriptsize,align=center] at (2.2,-1.25) {$\tau=10$: the slow eigenvalue is predicted well\\ in both cases, but at $d=0$ the full loop is unstable};
\end{scope}
\end{tikzpicture}}{Point 3. (a) A slow integrator $\tau\dot x_s=-k(x_2-x_2^\mathrm{ref})$ with $k=1/J_s$ acting on the two-state loop. (b) Its condensed model, whose slow eigenvalue is $-1/\tau$. (c) Eigenvalues of the full model for $\tau=10$. The condensed prediction of the slow eigenvalue is accurate in both cases, but with $d=0$ the slow layer destabilises the marginal fast loop. The condensed graph does not show this.}{fig:pt3}

\item \emph{Without attraction there is no stationary confounding.} When the noise excites the critical mode, $\|P\|$ and the confounding gain $\|\Pi\|$ diverge as the exit is approached from inside Regime~2b (Proposition~\ref{prop:exits}). Not every edge diverges in the same way: a pair of outputs in quadrature with the critical oscillation can keep a bounded instantaneous covariance while their cross-spectrum at the critical frequency diverges (Example~\ref{ex:hopf}).

\emph{Implication.} Covariance-based monitoring can miss an approaching exit if it watches the wrong pair. The frequency-resolved edges $\Phi(\omega)$ do not have this blind spot.

\ptfig{\begin{tikzpicture}[font=\small]
\begin{scope}
\node[font=\small\bfseries] at (-0.7,1.5) {(a)};
\node[nd] (p) at (0,0.4) {$x_1$}; \node[nd,minimum size=8mm,font=\scriptsize] (q) at (3.0,0.4) {$x_1{+}\tfrac12x_2$};
\draw[{Latex[length=2mm]}-{Latex[length=2mm]},thick,dashed] (p) to[bend left=25] node[lab,above]{\scriptsize $\kappa$: $0.625\to74.9$} (q);
\node[font=\scriptsize,align=center] at (1.5,-0.7) {in-phase pair: the edge weight\\ diverges like $1/d$};
\end{scope}
\begin{scope}[xshift=6.2cm]
\node[font=\small\bfseries] at (-0.7,1.5) {(b)};
\node[nd] (p2) at (0,0.4) {$x_1$}; \node[nd] (q2) at (3.0,0.4) {$x_2$};
\draw[{Latex[length=2mm]}-{Latex[length=2mm]},thick,dashed] (p2) to[bend left=25] node[lab,above]{\scriptsize $\kappa_{12}$: $-0.083\to-0.125$} (q2);
\draw[{Latex[length=2mm]}-{Latex[length=2mm]},thick,dashed] (p2) to[bend right=25] node[lab,below]{\scriptsize $|\Phi_{12}(\omega_0)|$: $0.48\to5300$} (q2);
\node[font=\scriptsize,align=center] at (1.5,-0.95) {quadrature pair: covariance bounded,\\ cross-spectrum diverges like $1/d^2$};
\end{scope}
\end{tikzpicture}}{Point 4, using the numbers of Example~\ref{ex:hopf} as $d$ falls from $1$ to $0.01$. (a) An in-phase pair: its bidirected edge grows without bound. (b) A quadrature pair: its instantaneous covariance stays bounded while its cross-spectrum at $\omega_0=\sqrt2$ diverges. Only the frequency-resolved edge sees the approaching exit.}{fig:pt4}

\item \emph{Equilibrium and dynamic semantics agree on the settled response, not on everything.} Inside Regime~2b both descriptions give the same means and the same effects of interventions on slow variables. The dynamic description then adds what the equilibrium model does not contain: the bidirected edges with weights $\kappa_{ij}=c_i\T Pc_j$, and their spectra.

\emph{Implication.} An equilibrium causal model of a feedback system is correct but incomplete in Regime~2b. The missing part is exactly the content of Section~\ref{sec:2b}.

\ptfig{\begin{tikzpicture}[font=\small]
\begin{scope}
\node[font=\small\bfseries] at (-0.7,1.6) {(a)};
\node[ex] (u) at (0,0.5) {$u$};
\node[nd] (y1) at (2.0,1.1) {$x_1$}; \node[nd] (y2) at (2.0,-0.1) {$x_2$};
\draw[dir] (u)-- node[lab,above]{\scriptsize $1/3$} (y1); \draw[dir] (u)-- node[lab,below]{\scriptsize $1/3$} (y2);
\node[font=\scriptsize,align=center] at (1.0,-1.1) {equilibrium semantics:\\ directed edges only};
\end{scope}
\begin{scope}[xshift=5.2cm]
\node[font=\small\bfseries] at (-0.7,1.6) {(b)};
\node[ex] (ub) at (0,0.5) {$u$};
\node[nd] (z1) at (2.0,1.1) {$x_1$}; \node[nd] (z2) at (2.0,-0.1) {$x_2$};
\draw[dir] (ub)-- node[lab,above]{\scriptsize $1/3$} (z1); \draw[dir] (ub)-- node[lab,below]{\scriptsize $1/3$} (z2);
\draw[{Latex[length=2mm]}-{Latex[length=2mm]},thick,dashed] (z1) to[bend left=40] node[lab,right]{\scriptsize $\kappa_{12}=-0.083$} (z2);
\node[font=\scriptsize,align=center] at (1.2,-1.1) {Regime~2b: same directed edges,\\ plus the induced bidirected edge};
\end{scope}
\end{tikzpicture}}{Point 5 for the two-state loop with $d=1$ and a constant input $u$ into $x_1$. (a) The equilibrium model has the directed edges $J_s=(1/3,1/3)$. (b) The Regime~2b model has the same directed edges and adds the bidirected edge with weight $\kappa_{12}=P_{12}=-0.083$ for unit noise on each state. The two semantics agree on (a) and differ only by the added edge.}{fig:pt5}

\item \emph{Persistence makes condensation robust, and the bounds make it quantitative.} With a time-scale gap, normal attraction gives persistence of the slow manifold under perturbation \cite{fenichel1979}. The quantitative statements are the certificate of Proposition~\ref{prop:cert}
and the separation needed for the noise, condition~\eqref{eq:noise-separation} of
Corollary~\ref{cor:noise-separation}.

\emph{Implication.} ``Enough separation'' is a number that can be computed. The noise usually needs more separation than the eigenvalues do, sometimes by an order of magnitude.

\ptfig{\begin{tikzpicture}[font=\small]
\begin{scope}
\node[font=\small\bfseries] at (-0.6,2.6) {(a)};
\draw[->] (0,0)--(4.2,0) node[right,font=\scriptsize]{$x_s$};
\draw[->] (0,0)--(0,2.2) node[above,font=\scriptsize]{$x_f$};
\draw[thick] plot[smooth,domain=0.2:4] (\x,{0.8+0.35*sin(deg(1.3*\x))});
\draw[thick,dashed] plot[smooth,domain=0.2:4] (\x,{0.95+0.33*sin(deg(1.3*\x+0.25))});
\foreach \xx in {0.8,1.9,3.0,3.8}{
  \draw[dir,thin] (\xx,{0.8+0.35*sin(deg(1.3*\xx))+0.75})--(\xx,{0.8+0.35*sin(deg(1.3*\xx))+0.2});
  \draw[dir,thin] (\xx,{0.8+0.35*sin(deg(1.3*\xx))-0.6})--(\xx,{0.8+0.35*sin(deg(1.3*\xx))-0.1});}
\node[font=\scriptsize,anchor=west] at (4.05,1.15) {$h(x_s)$};
\node[font=\scriptsize,align=center] at (2.0,-0.75) {the slow manifold attracts and persists\\ under perturbation (dashed)};
\end{scope}
\begin{scope}[xshift=10.6cm,yshift=0.9cm]
\node[font=\small\bfseries] at (-4.0,1.7) {(b)};
\draw[->] (-3.8,0)--(0.6,0) node[right,font=\scriptsize]{Re};
\draw[->] (0,-1.2)--(0,1.3) node[above,font=\scriptsize]{Im};
\fill (-3.0,0.8) circle (1.6pt); \fill (-3.0,-0.8) circle (1.6pt); \fill (-3.3,0) circle (1.6pt);
\draw[densely dotted] (-2.7,-1.1)--(-2.7,1.1) node[above,font=\scriptsize]{$\alpha$};
\draw (-0.45,0) circle (0.25); \fill (-0.48,0) circle (1.6pt);
\node[font=\scriptsize,anchor=south] at (-0.45,0.25) {$-1/\tau$};
\draw[<->] (-0.45,-0.42)--(-0.2,-0.42); \node[font=\scriptsize,anchor=north] at (-0.33,-0.45) {$\delta/\tau$};
\node[font=\scriptsize,align=center] at (-1.6,-1.55) {exactly one eigenvalue within $\delta/\tau$ of $-1/\tau$\\ when $\tau|\alpha|>4$ and $\delta<1$};
\end{scope}
\end{tikzpicture}}{Point 6. (a) Normal attraction: trajectories approach the slow manifold $x_f=h(x_s)$, and a perturbed manifold (dashed) exists nearby. (b) The quantitative counterpart, Propositions~\ref{prop:cert} and~\ref{prop:apriori}: with the fast eigenvalues to the left of $\alpha$, exactly one eigenvalue lies within $\delta/\tau$ of the condensed value $-1/\tau$. Schematic.}{fig:pt6}

\item \emph{Confounding is finite under stability alone, but finite is not small.} A Hurwitz $A$ is enough for the confounding gain $\|\Pi\|=\eta/(2|\alpha|)$ of \eqref{eq:gain} to be finite; the time-scale gap plays no part in finiteness. A fast block with large $|\alpha|$ can still confound strongly if it is strongly non-normal, because $\eta$ can be large.

\emph{Implication.} Making a loop faster does not by itself make its induced confounding negligible. The non-normality factor $\eta$ has to be checked as well.

\ptfig{\begin{tikzpicture}[font=\small]
\begin{scope}
\node[font=\small\bfseries] at (-0.7,1.6) {(a)};
\node[nd] (a1) at (0,0.5) {$x_1$}; \node[nd] (a2) at (2.0,0.5) {$x_2$};
\draw[dir] (a1) edge[loop above,looseness=6] node[lab,above]{\scriptsize $-1$} (a1);
\draw[dir] (a2) edge[loop above,looseness=6] node[lab,above]{\scriptsize $-1$} (a2);
\node[font=\scriptsize,align=center] at (1.0,-1.05) {$g=0$: $\alpha=-1$, $\eta=1$,\\ $\|\Pi\|=0.5$, $\kappa_{12}=0$};
\end{scope}
\begin{scope}[xshift=5.4cm]
\node[font=\small\bfseries] at (-0.7,1.6) {(b)};
\node[nd] (b1) at (0,0.5) {$x_1$}; \node[nd] (b2) at (2.0,0.5) {$x_2$};
\draw[dir] (b1) edge[loop above,looseness=6] node[lab,above]{\scriptsize $-1$} (b1);
\draw[dir] (b2) edge[loop above,looseness=6] node[lab,above]{\scriptsize $-1$} (b2);
\draw[dir,line width=1.4pt] (b2)-- node[lab,above]{\scriptsize $g=10$} (b1);
\draw[{Latex[length=2mm]}-{Latex[length=2mm]},thick,dashed] (b1) to[bend right=50] node[lab,below]{\scriptsize $\kappa_{12}=2.5$} (b2);
\node[font=\scriptsize,align=center] at (1.0,-1.05) {$g=10$: same $\alpha=-1$, $\eta=51.5$,\\ $\|\Pi\|=25.7$, $\operatorname{Var}x_1=25.5$};
\end{scope}
\end{tikzpicture}}{Point 7: two blocks with the same decay rate $\alpha=-1$ and unit noise on each state. (a) No coupling: $\eta=1$ and $\|\Pi\|=0.5$. (b) A one-way coupling $g=10$ makes the block non-normal: $\eta=51.5$, $\|\Pi\|=25.7$, the variance of $x_1$ rises from $0.5$ to $25.5$, and a bidirected edge of weight $2.5$ appears. Speed alone does not bound the confounding.}{fig:pt7}

\item \emph{Layers stack into a mixed graph, not a DAG.} When each layer is in Regime~2b and the layers are separated in time scale, the condensed layers form an acyclic directed mixed graph. Directed edges carry $J_s$; bidirected edges carry the induced confounding. A supervisory layer may plan interventions on the directed part. If it learns those edges from operating data and ignores the bidirected part, it can get even the sign of an effect wrong (Proposition~\ref{prop:id}, Example~\ref{ex:regression}).

\emph{Implication.} Planning uses $J_s$; learning needs instruments, injected excitation or explicit allowance for the bidirected edges.

\ptfig{\begin{tikzpicture}[font=\small]
\begin{scope}
\node[font=\small\bfseries] at (-0.7,2.5) {(a)};
\node[ex] (z) at (0,0.6) {$z$};
\node[nd] (w1) at (1.6,1.3) {$w_1$}; \node[nd] (w2) at (1.6,-0.1) {$w_2$};
\node[nd] (y1) at (3.3,1.3) {$y_1$}; \node[nd] (y2) at (3.3,-0.1) {$y_2$};
\draw[dir] (z)--(w1); \draw[dir] (z)--(w2); \draw[dir] (w1)--(y1); \draw[dir] (w2)--(y2); \draw[dir] (w1)--(y2);
\draw[{Latex[length=2mm]}-{Latex[length=2mm]},thick,dashed] (w1) to[bend right=40] (w2);
\draw[{Latex[length=2mm]}-{Latex[length=2mm]},thick,dashed] (y1) to[bend left=40] (y2);
\node[font=\scriptsize] at (0,1.9) {supervisor}; \node[font=\scriptsize] at (1.6,1.95) {middle}; \node[font=\scriptsize] at (3.3,1.95) {fast};
\node[font=\scriptsize,align=center] at (1.7,-1.1) {stacked layers: an acyclic\\ directed mixed graph};
\end{scope}
\begin{scope}[xshift=6.2cm]
\node[font=\small\bfseries] at (-0.7,2.5) {(b)};
\node[ex] (r) at (0,0.6) {$r$};
\node[nd] (x1) at (1.5,0.6) {$x_1$}; \node[nd] (x2) at (3.2,0.6) {$x_2$};
\draw[dir] (r)--(x1); \draw[dir] (x1)-- node[lab,below]{\scriptsize $+1$} (x2);
\draw[{Latex[length=2mm]}-{Latex[length=2mm]},thick,dashed] (x1) to[bend left=45] (x2);
\node[font=\scriptsize,align=left,anchor=west] at (4.0,0.6) {least squares: $-0.125$\\ instrument $r$: $0.990$};
\node[font=\scriptsize,align=center] at (2.2,-1.1) {learning an edge from operating data:\\ the bidirected edge biases least squares};
\end{scope}
\end{tikzpicture}}{Point 8. (a) Layers in Regime~2b with separated time scales stack into an acyclic directed mixed graph: directed edges between layers, and bidirected edges within each layer. A supervisor $z$ plans on the directed edges. (b) Learning the edge $x_1\to x_2$ of the two-state loop ($d=1$, true effect $+1$) from operating data: least squares returns $-0.125$ because of the bidirected edge, while an injected excitation $r$ used as an instrument returns $0.990$ (Example~\ref{ex:regression}).}{fig:pt8}
\end{enumerate}

Taken together, the eight points give the following summary, which is the form in which the distinction should be quoted. A cycle that is uniquely solvable but not attracting (Regime~2a without 2b) has a well-defined equilibrium that the system does not settle to. Its condensed model gives the correct shift of that equilibrium under interventions that leave the loop intact, but the system does not reach the predicted state. A slower layer has no grounds to treat the cycle as condensed and, when noise excites the critical mode, the cycle has no stationary confounding structure. Normal attraction (Regime~2b) closes this gap. The equilibrium semantics gives the settled response, and the dynamic semantics adds finite, computable bidirected edges with confounding gain $\eta/(2|\alpha|)$. With a time-scale gap, condensation persists under perturbation, in the sense of Fenichel's theory and the finite-separation bounds. Layers then stack into an acyclic directed mixed graph. Supervisory layers can plan on its directed edges, but any edge learned from operating data must allow for the bidirected ones.

\section{What the formulation offers, in plain terms}\label{sec:offers}

Most engineered systems are built in layers. Fast loops hold currents, speeds, pressures or positions, and slower layers decide what those loops should aim for, estimate what is going on, or optimise. The framework gives a designer of such systems a few concrete things.

\emph{A test of what the slower layers may assume.} A slower layer usually treats the fast loops as if they had already settled. Regime~2a says when that settled picture exists at all. Regime~2b says when it is also stable enough to describe the fast loops' noise, and Section~\ref{sec:separation} says how much slower the upper layer must be for the picture to be accurate, separately for its dynamics and for its noise. The noise often needs far more separation: in a vehicle platoon the supervisor's dynamics were fine with a ratio of 5, but its noise needed about 76.

\emph{A warning that arrives before the trouble.} When a loop approaches a Hopf, flip or Neimark--Sacker boundary, its settled behaviour does not change, so models built on equilibria see nothing. Its fluctuations do change: they grow, and outputs start moving together. Watching the induced edges, or the confounding gain, gives early warning of the loss of stability.

\emph{One number to compare designs, with two knobs inside it.} The confounding gain says how strongly a loop spreads its disturbances into the signals around it. Its two factors say what to fix. A small $|\alpha|$ means the loop is too slow to forget disturbances near some operating condition; in a power converter on a weak grid this was the problem, and the cure was a slower, better damped phase-locked loop. A large $\eta$ means disturbances are being passed along and amplified between states, as in a chain of vehicles; there the cure was a change of topology, not of speed.

\emph{A clean split of the tuning job.} With integral action or an internal model, where the loop settles is fixed by the plant and the reference, whatever the gains. The gains then only decide how tightly the loop holds that point and how much noise it spreads. A designer can treat these as two separate questions, and an upper layer can be designed against the settled map before the lower layer is tuned.

\emph{Correct reading of data.} Data recorded while loops are running are confounded by the loops. A regression can get even the sign of an effect wrong (Example~\ref{ex:regression}). The framework says when this happens, by how much, and what fixes it: an injected test signal that the loop's own noise does not touch. It also says how much information a set of sensors really carries when they share a hidden cause, which is often far less than their sensor noise suggests.

\emph{Interfaces between layers.} A layer that reads a regulated signal learns nothing at the frequencies the regulation covers; it should read the control effort instead. A layer whose channel is delayed or quantised has a best update rate that no fixed rule of thumb gives. These are statements about what layers should exchange and how fast, which is the question that theories of layered control architectures \cite{matni2024quantitative} put at the centre.

The same steps apply to a converter, a drive, a vehicle platoon, a process plant, a building, a digital controller or a power network: find the fast block and what reads it, check the two regimes, weigh the induced edges, and choose the rates and interfaces of the layers above. Section~\ref{sec:use} gives these steps in checklist form.

\section{Worked examples}\label{sec:examples}

All numbers in this section were computed with the scripts that accompany the paper.

\begin{figure}[!htbp]
\centering
\includegraphics[width=\textwidth]{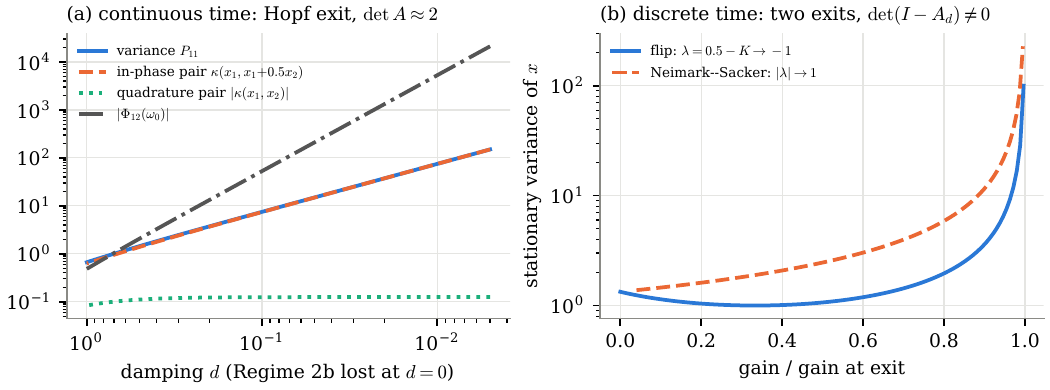}
\caption{Leaving Regime~2b while staying in Regime~2a. (a) The two-state loop of Example~\ref{ex:hopf} as its damping $d$ goes to zero (right): variances and in-phase covariances grow like $1/d$, the quadrature covariance stays bounded, and the cross-spectrum at the critical frequency grows like $1/d^2$. (b) The two digital loops of Example~\ref{ex:discrete}: stationary variance against gain, normalised by the gain at the exit.}
\label{fig:exits}
\end{figure}

\begin{example}[Hopf exit of a two-state loop]\label{ex:hopf}
Take the two-state loop \eqref{eq:two1} with $a=-2$, $b=1$ and unit noise. Its eigenvalues are $-d\pm j\sqrt2$, and $\det A=d^2+2$, so it is in Regime~2a for every $d$ and in Regime~2b for $d>0$. As $d$ falls from $1$ to $0.01$, the variance $P_{11}$ grows from $0.667$ to $75.0$, like $1/d$. The induced covariance between the in-phase pair $x_1$ and $x_1+0.5x_2$ grows from $0.625$ to $74.9$, with correlation between $0.92$ and $0.94$ throughout. The covariance between $x_1$ and $x_2$ stays bounded, moving from $-0.083$ to $-0.125$, because the two states are in quadrature, but the cross-spectrum $|\Phi_{12}(\omega_0)|$ at $\omega_0=\sqrt2$ grows from $0.48$ to $5300$, like $1/d^2$ (Figure~\ref{fig:exits}(a)). Throughout, the condensed map $-A^{-1}H$ changes smoothly. By contrast, along the fold route with $a=b$ and $ab$ rising towards $d^2=1$, the covariance $P_{12}$ grows from $0.71$ at $ab=0.5$ to $49.7$ at $ab=0.99$ while $\det A$ falls to $0.01$; there Regime~2a fails as well.
\end{example}

\begin{example}[Non-normal amplification in a feedforward chain]\label{ex:chain}
Take $N$ identical stages, $\dot x_1=-x_1+\dot w_1$ and $\dot x_i=-x_i+cx_{i-1}+\dot w_i$, with identity metric. The eigenvalues are all $-1$, so $\alpha=-1$ for every $N$ and $c$, and a design rule based on eigenvalues sees no difference between the cases. The confounding gain does. For $c=1$, $\eta$ is $1.81$, $3.73$ and $8.15$ at $N=2$, $4$ and $8$. For $c=2$, it is $3.41$, $37.4$ and $6230$, and the variance of the last stage grows from $1.5$ to $2354$. All of the growth is in $\eta$. The block is a directed acyclic graph of one-state loops, and its confounding comes entirely from the forward couplings. The same structure, a chain of loops, drives string instability in vehicle platoons \cite{seiler2004,singh2026c}. Sampling this chain with $c=2$, $N=3$ at $h=0.1$, $0.01$ and $0.001$ reproduces the continuous covariance through the Stein equation \eqref{eq:stein} to within $3\times10^{-14}$, as Proposition~\ref{prop:sampling} requires.
\end{example}

\begin{example}[Two discrete-time exits]\label{ex:discrete}
A scalar digital loop $x(k+1)=0.5x(k)+u(k)+w(k)$ with $u=-Kx$ has closed-loop pole $0.5-K$ and $1-A_d=0.5+K\ne0$ for $K>0$: it is in Regime~2a for all such gains. As $K$ rises towards $1.5$ the pole approaches $-1$, a flip. The stationary variance of $x$ grows from $1.33$ at $K=1$ to $50.3$ at $K=1.49$, and the induced covariance between the measured state and the control signal, $-K\operatorname{Var}x$, from $-1.33$ to $-74.9$. Adding integral action, $u=g\xi$ with $\xi(k+1)=\xi(k)-x(k)$, gives $\det(I-A_d)=g\ne0$ and a complex pair whose modulus reaches $1$ at $g=0.5$, a Neimark--Sacker crossing; the variance of $x$ grows from $1.61$ at $g=0.1$ to $57.3$ at $g=0.49$ (Figure~\ref{fig:exits}(b)). Neither loop can be detected from its equilibrium map. The flip has no continuous-time counterpart in a sampled system, so it is specific to genuinely discrete-time designs.
\end{example}

\begin{example}[Least squares reverses the sign of an effect]\label{ex:regression}
Take the two-state loop \eqref{eq:two1} with $d=1$, $a=-2$, $b=1$. Clamping $x_1$ at $v$ moves the equilibrium of $x_2$ to $(b/d)v$, so the causal effect of $x_1$ on $x_2$ in the condensed model is $+1$. A regression of $x_2$ on $x_1$ from ambient data gives $P_{12}/P_{11}=d(a+b)/(a^2-ab+2d^2)=-0.125$: the wrong sign, because the feedback $x_2\to x_1$ and the shared noise make the loop a hidden common cause of both. Injecting a slow excitation $r$ into $x_1$ only (an Ornstein--Uhlenbeck process with rate $0.01$) changes the least-squares slope to $+0.035$, still far off. The instrumental-variable estimate $\operatorname{Cov}(x_2,r)/\operatorname{Cov}(x_1,r)$ is $b/(d+0.01)=0.990$ exactly, and a Monte Carlo run of $10^5$ time units gives $0.973$ for the instrument and $0.035$ for least squares. This is Proposition~\ref{prop:id} at work.
\end{example}

\begin{example}[Integral action and an uninformative signal]\label{ex:pi}
In the PI loop of Figure~\ref{fig:pi}, the equilibrium ordering is $(r,n_e)\to x_p\to u\to\xi$, computed by bipartite matching on the equilibrium equations. Any layer above it that reads the measured regulated signal sees $r$ at zero frequency, whatever the plant does (Corollary~\ref{cor:regulated}). It should read $u$, which carries the disturbance, and whose zero-frequency value does not depend on the PI gains (Proposition~\ref{prop:gainfree}).
\end{example}

\subsection{Two applications}
\emph{A grid-following converter} \cite{singh2026a,singh2026b}. A twelve-state averaged model with current control, phase-locked loop and control delay leaves Regime~2b through a Hopf crossing near short-circuit ratio 1.46, through an oscillatory mode at about 19\,Hz, while its condensed map is unchanged. Its confounding is $\alpha$-dominated: $\eta$ stays between about 4 and 6 while $1/(2|\alpha|)$ diverges. Ambient admittance estimates are biased by 30--60\% on weak grids, and a redesign that keeps $|\alpha|$ near $22\,\mathrm{s}^{-1}$ lowers the worst-case confounding gain fivefold.

\emph{A vehicle platoon} \cite{singh2026c}. The fast block is a chain of per-vehicle loops, as in Example~\ref{ex:chain}, and its confounding is $\eta$-dominated: the decay rate is flat in platoon length while $\eta$ grows to about $255$ for predecessor following and $111$ for cooperative adaptive cruise control at sixteen vehicles. A supervisor that estimates a common load from the vehicles' control efforts overstates its information a hundredfold if it ignores the induced spectrum, and it needs a time-scale ratio of about $76$ for 10\% accuracy in its variance.

\section{Using the framework}\label{sec:use}

For later work the framework reduces to a short sequence of checks.

\begin{enumerate}[label=\arabic*.]
\item \emph{Identify the fast block and what reads it.} Write the block in the form \eqref{eq:ct} or \eqref{eq:dt}, list the outputs read by slower layers or analysts, and compute the dynamic causal graph and the equilibrium causal ordering.
\item \emph{Check Regime~2a.} Is $A$ (or $I-A_d$) nonsingular over the operating range? If so, compute $J_s$; under integral action or an internal model it does not depend on the controller gains.
\item \emph{Check Regime~2b.} Is $A$ Hurwitz (or $A_d$ Schur) over the operating range? Locate the nearest exit and its type: Hopf, flip or Neimark--Sacker leave the condensed map intact and make the edges diverge.
\item \emph{Weigh the edges.} Solve \eqref{eq:lyap} or \eqref{eq:stein} for $P$, compute $\kappa$ and $\Phi(\omega)$, and compute $\|\Pi\|$ in a declared metric with its split into $1/(2|\alpha|)$ and $\eta$.
\item \emph{Separate the layers.} For each slower layer, check the eigenvalue certificate
(Proposition~\ref{prop:cert}) and the noise condition~\eqref{eq:noise-separation}
for a declared tolerance $\varepsilon_{\mathrm{tol}}$.   
\item \emph{Plan identification.} Wherever a directed effect is to be estimated from closed-loop data, provide an excitation uncorrelated with the block's own noise.
\end{enumerate}

\noindent We suggest the following conventions for subsequent work: \emph{Regime~2a} and \emph{Regime~2b} as in Definitions~\ref{def:2a} and~\ref{def:2b}; \emph{condensation} for the replacement of a Regime~2a block by its equilibrium map; \emph{induced confounding} for $\kappa$ and $\Phi$; \emph{confounding gain} for $\|\Pi\|$ with $\eta$ and $1/(2|\alpha|)$ as its \emph{inflation} and \emph{slowing} factors; and \emph{defect} for $E_k$. Table~\ref{tab:notation} collects the notation.

\begin{table}[!htbp]
\centering\small
\caption{Notation.}
\label{tab:notation}
\begin{tabular}{>{\raggedright\arraybackslash}p{0.22\textwidth}>{\raggedright\arraybackslash}p{0.33\textwidth}>{\raggedright\arraybackslash}p{0.33\textwidth}}
\toprule
 & Continuous time & Discrete time\\
\midrule
Fast block & $\dot x_f=Ax_f+Hx_s+B\dot w$ & $x_f(k{+}1)=A_dx_f(k)+H_dx_s(k)+w(k)$\\
Regime 2a & $\det A\ne0$ & $\det(I-A_d)\ne0$\\
Condensed sensitivity & $J_s=D_p-CA^{-1}H$ & $J_s=D_p+C(I-A_d)^{-1}H_d$\\
Regime 2b & $A$ Hurwitz, $\alpha<0$ & $A_d$ Schur, $\rho<1$\\
Stationary covariance & $AP+PA\T+BB\T=0$ & $P=A_dPA_d\T+Q_d$\\
Induced confounding & $\kappa_{ij}=c_i\T Pc_j$, $\Phi(\omega)$ & $\kappa_{ij}=c_i\T Pc_j$, $\Phi(\theta)$\\
Unit-noise Gramian & $\tilde A\Pi+\Pi\tilde A\T+I=0$ & $\Pi_d=\tilde A_d\Pi_d\tilde A_d\T+I$\\
Confounding gain & $\|\Pi\|=\eta/(2|\alpha|)$ & $\|\Pi_d\|=\eta_d/(1-\rho^2)$\\
Defect & $E_k=\tilde B\tilde B\T-2kR$ & $E_k=\tilde Q_d-k(I-\tilde A_d\tilde A_d\T)$\\
Exits from 2b keeping 2a & Hopf & flip, Neimark--Sacker\\
\bottomrule
\end{tabular}
\end{table}

\section{Scope and open problems}\label{sec:scope}

The results are linear, or linearised about an operating point, with white noise. In the nonlinear case the natural setting is that of NAIMs (Section~\ref{sec:naim}): the slow manifold $x_f=h(x_s)$ replaces $-A^{-1}Hx_s$, Regime~2a becomes existence and local uniqueness of that manifold (a nonsingular normal linearisation) and Regime~2b becomes its normal attraction \cite{fenichel1979,kokotovic1986}, and the confounding gain becomes a function on the manifold, with a contraction metric \cite{lohmiller1998} in place of the storage metric. The noise results then hold for small noise away from loss of normal attraction. Near a Hopf or flip exit, noise drives large excursions and metastability, and genuinely nonlinear tools are needed. Coloured and non-stationary noise, vector-valued slow layers with interacting rates, learning the block from data in the presence of its own induced edges, and a full theory of interventions on fast variables that change the regime are open. The relation between Regime~2b and the dynamic structural causal models of \cite{bongers2018,boeken2024} deserves a careful treatment of its own.

\section*{Reproducibility}
The examples were computed in Python with NumPy and SciPy; the closed forms \eqref{eq:two2} were checked symbolically with SymPy, and the causal orderings of Sections~\ref{sec:notsink} and~\ref{sec:manystate} were computed with NetworkX. The scripts accompany this paper.
\section*{Acknowledgment}
The authors used Claude (Anthropic) for editorial and verification
assistance during the preparation of this manuscript. This included
checking in-text citations; checking the
numbering of equations, propositions and cross-references; comparing
successive versions of the manuscript and suggesting references for statements that lacked citations. The theoretical framework, results,
examples, figures and simulations are the authors' own. All suggested material was reviewed,
verified and edited by the authors, who take full responsibility for
the content of this publication.

\end{document}